\documentclass[]{article}

\usepackage{amsmath}
\usepackage{amsfonts}
\usepackage{url}

\usepackage[margin=0.8in]{geometry}
\usepackage{graphicx}
\usepackage{caption}
\usepackage[labelformat=parens,labelfont={}]{subcaption}
\usepackage{amsthm}
\newtheorem*{remark}{Remark}

\newcommand{\bs}{\boldsymbol}

\newcommand{\norm}[1]{\left|\!\left| #1 \right|\!\right|}
\newcommand{\ol}{\overline}

\newcommand{\dif}{\hspace{0.1cm}\mathrm{d}}
\newcommand{\D}{\mathrm{D}}

\newcommand{\on}{\text{on }}

\makeatletter
\newcommand*{\rmnum}[1]{\expandafter\@slowromancap\romannumeral #1@}
\makeatother

\begin{document}
\title{Unconditionally stable, second-order accurate schemes for solid state phase transformations driven by mechano-chemical spinodal decomposition}
\author{K. Sagiyama\thanks{Department of Mechanical Engineering, University of Michigan}, S. Rudraraju\thanks{Department of Mechanical Engineering, University of Michigan} \& K. Garikipati\thanks{Departments of Mechanical Engineering, and Mathematics, University of Michigan, corresponding author, {\tt krishna@umich.edu}}}
\maketitle
\abstract{
We consider solid state phase transformations that are caused by free energy densities with domains of non-convexity in strain-composition space; we refer to the non-convex domains as mechano-chemical spinodals.  
The non-convexity with respect to composition and strain causes segregation into phases with different crystal structures.  
We work on an existing model that couples the classical Cahn-Hilliard model with Toupin's theory of gradient elasticity at finite strains. Both systems are represented by fourth-order, nonlinear, partial differential equations.  
The goal of this work is to develop unconditionally stable, second-order accurate time-integration schemes, motivated by the need to carry out large scale computations of dynamically evolving microstructures in three dimensions.  
We also introduce \emph{reduced} formulations naturally derived from these proposed schemes for faster computations that are still second-order accurate.  
Although our method is developed and analyzed here for a specific class of mechano-chemical problems, one can readily apply the same method to develop unconditionally stable, second-order accurate schemes for any problems for which free energy density functions are multivariate polynomials of solution components and component gradients.  
Apart from an analysis and construction of methods, we present a suite of numerical results that demonstrate the schemes in action.  
}
\section{Introduction}\label{S:introduction}
Many multicomponent solids undergo phase-transformations in which a diffusional redistribution of their different components is coupled with a structural change of the crystallographic unit cell.  
One example is the phase-transformation of yttria-stabilized zirconia $\textrm{Zr}_{1-x}\textrm{Y}_x\textrm{O}_{2-x/2}$ from cubic at high-Y composition to tetragonal at low-Y if quenched to a low temperature. Another is observed in the spinel structures of Li-ion electrodes when cubic $\textrm{LiMnO}_4$ transforms into tetragonal $\textrm{Li}_2\textrm{Mn}_2\textrm{O}_4$ upon discharging to low voltages. Pure $\textrm{ZrO}_2$ and $\textrm{TaS}_2$ are other materials that may be susceptible to such mechano-chemical phase transformations.

The underlying phenomenology can be described by a free energy density function, which is non-convex in its mechanical (strain) and chemical (composition) arguments. Cahn and Hilliard \cite{CahnHilliard1958} famously introduced the chemical spinodal as the domain in composition space where the free energy density is non-convex, and varies smoothly between local minima that correspond to distinct phases. 
This notion has recently been extended to the mechano-chemical spinodal, defined as the domain in strain-composition space where the Hessian of the (sufficiently smooth) free energy density function has non-positive eigenvalues \cite{Rudrarajuetal2015}; see Fig. \ref{Fi:plot_Psic_Psie_3D}. 
If the state of a material point lies within the mechano-chemical spinodal, and specifically if the free energy density is non-convex with respect to composition, the solid will undergo diffusional segregation. 
Here we concern ourselves with cases in which the two resulting phases have cubic and tetragonal crystal structures, respectively. 
The cubic structure corresponds to a minimum of the free energy density function in strain-composition space. 
Furthermore, we adopt the undistorted cubic structure as the reference state for strain. 
Then, the tetragonal lattice is naturally obtained by the strain relative to the cubic structure. 
The symmetries of the cubic lattice split into three identical sub-groups, each of which corresponds to a tetragonal lattice oriented along one of the cubic crystal axes. 
These three tetragonal variants, however, correspond to different strains relative to the reference cubic lattice. 
The free energy density function admits three additional minima, each corresponding to the strain that transforms the reference cubic lattice into one of the tetragonal variants. 
However, the compositions at these additional minima are identical. 
As the states of material points traverse such a multi-welled free energy density surface in strain-composition space the solid develops a microstructure. 
The mechano-chemical spinodal is regarded as a domain of instability since small fluctuations in strain and composition tend to grow as the state evolves towards one of the wells in strain-composition space. Stress softening and ``uphill'' diffusion result, respectively.
  
A non-convex free energy density function can lead to microstructure as explained above. 
However, a mathematical model restricted to the above phenomenology leads to ill-posed partial differential equations (PDEs) for elasticity and transport characterized by spurious mesh dependence. 
The volume fractions of the cubic and tetragonal phases would be set by initial and boundary conditions on the transport problem, and the microstructural pattern of tetragonal variants would depend on mechanical boundary conditions. 
However, in the absence of intrinsic length scales in the mathematical model, the extent and thickness of interfaces between phases and variants would be determined by the mesh. 
There is a well-understood physical aspect to this argument, also: The model promotes free energy minimizing microstructures, but incurs no penalty for the strain and composition gradients as these fields fluctuate between one phase or variant and the next. 
Arbitrarily fine energy minimizing microstructures are therefore admissible---an essentially unphysical result. 
Mathematical well-posedness and physical realism are restored by extending the free energy density function to include a dependence on gradient fields of strain and composition. 
The corresponding free energy coefficients introduce intrinsic length scales, and the gradient energies distinguish between microstructures of differing fineness, penalizing those that vary rapidly. 

Gradient free energies in classical settings lead to the Cahn-Hilliard equation for mass transport \cite{CahnHilliard1958}, and variants of strain gradient elasticity that represent size effects. 
Because the transformation strains between the cubic/tetragonal phases and between the tetragonal variants are of finite magnitude in many material systems, we are led to Toupin's theory of nonlinear (finite strain) gradient elasticity \cite{Toupin1964}.  
The Cahn-Hilliard equation and Toupin's strain gradient elasticity present fourth-order spatial derivatives in primal strong form, with corresponding weak forms carrying second-order derivatives.  
Carrying out large scale computations of these dynamically evolving microstructures in three dimensions is challenging; efficient and accurate time-integration algorithms are demanded.  
Rudraraju and co-workers used the Backward Euler algorithm \cite{Rudrarajuetal2015} to solve the same problem that we consider here.  
However, in that work the authors concerned themselves with introducing the notion of the mechano-chemical spinodal, and exploring the associated physics; not with a development/an analysis of accurate schemes, which is the goal of the present communication. 

Many unconditionally stable schemes have been proposed for the Cahn-Hilliard and related equations.  
The key ingredient of developing unconditionally stable schemes is appropriate temporal approximation of the chemical potential term, or the first derivative of the non-convex chemical free energy density with respect to chemical composition.  
Most often used are \emph{convex splitting} methods \cite{Elliott1993,Eyre1998}, where concave and convex parts of the chemical free energy are treated by different approximation methods.  
Convex splitting methods have been used in various related equations such as phase-field crystal equation \cite{Wise2009,Calo2014}, the Cahn-Hilliard-Hele-Shaw system \cite{Wise2010}, multicomponent Cahn-Hilliard equations \cite{Boyer2011,Tavakoli2016}, and the Navier-Stokes-Cahn-Hilliard equation \cite{Gao2012,Han2015}.  
The \emph{midpoint approximation} method introduced in \cite{Elliott1989a} is an alternative that does not introduce numerical dissipation and it has been used for the Cahn-Hilliard equation \cite{Du1991}, the Allen-Cahn equation and the Cahn-Hilliard equation \cite{GuillenGonzalez2014}, the Navier-Stokes-Cahn-Hilliard equation \cite{Feng2006,Hua2011}, liquid crystal dynamics \cite{Lin2007}, and a two-phase flow model \cite{Hyon2010}.  
The truncated Taylor expansion method was proposed for the Cahn-Hilliard equation in \cite{Kim2004}.  
The advantage of this method is that it can be extended to obtain unconditionally stable schemes for multicomponent Cahn-Hilliard equations \cite{Kim2004a} using truncated multivariate Taylor expansion.  
In these formulations remainder terms of the Taylor polynomials serve as numerical dissipation; indeed, if \emph{untruncated} Taylor expansion were used for the standard single-variate non-convex quartic chemical free energy density, the resulting approximation would coincide with the midpoint approximation that has no numerical dissipation.  
Truncated Taylor expansion is also used in conjunction with convex splitting methods in \cite{Wu2014} for the Cahn-Hilliard equation.  
Note that the composition gradient appearing in the Cahn-Hilliard equation is separately approximated in the above midpoint approximation method and Taylor expansion methods, but is treated as a unified contribution in convex splitting methods.  
Most of the above works consider quartic chemical free energy density.  
In \cite{Gomez2011} unconditionally stable schemes were developed for the Cahn-Hilliard equation for logarithmic free energy densities, using dedicated quadrature formula.  

The mechano-chemical free energy densities that we focus on in this work can be separated into two parts; viz.\ a logarithmic chemical free energy density used in \cite{Gomez2011} and a non-convex multivariate polynomial function that is eighth-order in strain and second-order in chemical composition, composition gradient, and strain gradient.  
Since the coupling of these terms makes the non-convex contribution rather complex, it is beneficial to develop a method that deals with components and component gradients in a unified framework instead of treating each of them separately.  
Especially, a framework that provides accurate schemes in the presence of strain and strain gradients is crucial.  
To this end, we employ multivariate Taylor expansions as in \cite{Kim2004a}, but we now regard composition gradients and strain gradients as direct variables in addition to the composition and strain themselves, and use \emph{untruncated} multivariate Taylor expansions to ensure unconditional stability and absence of numerical dissipation.  
Finally, the logarithmic chemical free energy density is treated in a similar way using the Taylor expansion, partly drawing from the analysis in \cite{Gomez2011}.  
The resulting scheme also enjoys second-order accuracy as those proposed in \cite{Gomez2011,Kim2004a}.  
To our knowledge this is the first unconditionally stable time-integration algorithm for mechano-chemical phase-transformation problems. 
Of note is the incorporation of gradient elasticity at finite strains---a feature that adds considerable complexity to the equations.  
The proposed method can be applied to any free energy density functions that are multivariate polynomials of components, component gradients, and higher-order component gradients to obtain unconditionally stable, second-order schemes.  
Finally, in addition to that they are accurate, these formulations allow for a straightforward simplification to produce practically useful second-order schemes that require less computation; those simplified formulations are called \emph{reduced} formulations and are also investigated in this work.  

In Sec. \ref{S:variational_formulation} we present the variational formulation of our mechano-chemical problem.  
The corresponding spatially and temporally discrete formulations are developed in Sec. \ref{S:numerical_formulation}.  
Unconditional stability and second-order accuracy of our fully discrete formulation are studied in Sec. \ref{S:analysis}.  
Finally, a suite of numerical examples is presented in Sec. \ref{S:numerical_example} to demonstrate the performance of the algorithms in three dimensions.  
Final remarks are made and future work proposed in Sec. \ref{S:conclusion}.  

\section{Variational formulations for mechano-chemical spinodal decomposition}\label{S:variational_formulation}
We consider mechano-chemical spinodal decomposition in a body that occupies, at the initial time, a bounded domain $\Omega$ in three-dimensional Euclidean space, in which we introduce the rectangular Cartesian coordinate system with $X_J$ ($J=1,2,3$) the corresponding coordinate variables.  

We are interested in the chemical composition field $c(\bs{X},t)\in(0,1)$ and the mechanical displacement field $\bs{u}(\bs{X},t)$ in $\Omega$.  
In this section we assume that these quantities and their spatial derivatives are continuously defined in $\ol{\Omega}$.  
The boundary of $\Omega$ is assumed to be decomposed into a finite number of smooth surfaces $\Gamma_{\iota}$, smooth curves $\Upsilon_{\iota}$, and points $\Xi_{\iota}$, so that $\partial\Omega=\Gamma\cup\Upsilon\cup\Xi$ where $\Gamma=\cup_{\iota}\Gamma_{\iota}$, $\Upsilon=\cup_{\iota}\Upsilon_{\iota}$, and $\Xi=\cup_{\iota}\Xi_{\iota}$.  
Each surface $\Gamma_{\iota}$ and curve $\Upsilon_{\iota}$ is further divided into mutually exclusive Dirichlet and Neumann subsets that are represented, respectively, by superscripts of lowercase letters $u$, $m$, and $g$ and those of uppercase letters $T$, $M$, and $G$, as $\Gamma_{\iota}=\Gamma_{\iota}^u\cup\Gamma_{\iota}^T=\Gamma_{\iota}^m\cup\Gamma_{\iota}^M$ and $\Upsilon_{\iota}=\Upsilon_{\iota}^g\cup\Upsilon_{\iota}^G$.  
We also denote by $\Gamma^u=\cup_{\iota}\Gamma_{\iota}^u$, $\Gamma^T=\cup_{\iota}\Gamma_{\iota}^T$, $\Gamma^m=\cup_{\iota}\Gamma_{\iota}^m$, $\Gamma^M=\cup_{\iota}\Gamma_{\iota}^M$,  $\Upsilon^g=\cup_{\iota}\Upsilon_{\iota}^g$, and $\Upsilon^G=\cup_{\iota}\Upsilon_{\iota}^G$ the unions of the Dirichlet and Neumann boundaries.  
Our formulation and analysis presented in the following can be readily extended to include mixed boundary conditions.  
As in \cite{Toupin1964}, coordinate derivatives of a scalar function $\phi$ are decomposed on $\Gamma$ into normal and tangential components as:
\begin{align}
\phi_{,J}=D\phi N_J+D_J\phi,\notag
\end{align}
where
\begin{align}
D\phi   &:=\phi_{,K}N_K,\notag\\
D_J\phi &:=\phi_{,J}-\phi_{,K}N_KN_J,\notag
\end{align}
where $N_J$ are the components of the unit outward normal to $\Gamma$.  
Here as elsewhere ${(\hspace{1pt}\cdot\hspace{1pt})_{,J}}$ denotes the spatial derivative with respect to the reference coordinate variable $X_J$.  

Dirichlet boundary conditions for the displacement field $\bs{u}$ can now be given as:
\begin{align}
u_i=\bar{u}_i\quad\on\Gamma^u,\quad
Du_i=\bar{m}_i\quad\on\Gamma^m,\quad
u_i=\bar{g}_i\quad\on\Upsilon^g,\label{E:u_DirichletBC}
\end{align}
where $\bar{u}_i$, $\bar{m}_i$, and $\bar{g}_i$ are components of known vector functions on $\Gamma^u$, $\Gamma^m$, and $\Upsilon^g$.  
On the other hand, we denote the components of the standard surface traction on $\Gamma^T$, the higher-order traction on $\Gamma^M$, and the line traction on $\Upsilon^G$ by $\bar{T}_i$, $\bar{M}_i$, and $\bar{G}_i$, whose mathematical formulas will be clarified in Sec.\ref{SS:variational_formulation_mechanics}.  
The chemical composition field $c$ is assumed to have no Dirichlet boundary conditions throughout $\partial\Omega$.  

    \subsection{Free energy}\label{SS:variational_formulation_free-energy}
    We derive the initial and boundary value problem for mechano-chemical spinodal decomposition guided by variational consideration.  
The total free energy that we consider in this work is a functional of $c$ and $\bs{u}$ defined as:
\begin{align}
\Pi\left[c,\bs{u}\right]
:=\int_{\Omega}
\Psi_c+\Psi_s+\Psi_e
\dif V
-\int_{\Gamma^T}u_i\bar{T}_i\dif S
-\int_{\Gamma^M}Du_i\bar{M}_i\dif S
-\int_{\Upsilon^G}u_i\bar{G}_i\dif C,
\label{E:Pi}
\end{align}
where $\Psi_c(c)$, $\Psi_s(c,c_{,A})$, and $\Psi_e(c,F_{iJ},F_{iJ,K})$ are the chemical, surface, and mechanical free energy densities that are functions of chemical composition, $c$, gradient of chemical composition, $c_{,A}$, deformation gradient, $F_{iJ}$, and gradient of deformation gradient, $F_{iJ,K}$, at each fixed point $\bs{X}\in\Omega$. Here, $F_{iJ}=\delta_{iJ}+u_{i,J}$ are the components of the deformation gradient tensor.  
These free-energy densities are explicitly defined as:
\begin{subequations}
\begin{align}
\Psi_c&:=A_1\left(c\log c+\left(1-c\right)\log\left(1-c\right)\right)+A_2\phantom{\cdot}c\left(1-c\right),\label{E:Psi_c}\\
\Psi_s&:=\frac{1}{2}c_{,A}K_{AB}(c)\hspace{1pt}c_{,B},\label{E:Psi_s}\\
\Psi_e&:
=B_1(c)\left(e_1-e_{\text{chem}}\left(c\right)\right)^2\notag\\
&\phantom{:}+B_2(c)\left(e_2^2+e_3^2\right)
+B_3(c)e_3\left(e_3^2-3e_2^2\right)
+B_4(c)\left(e_2^2+e_3^2\right)^2
+B_5(c)\left(e_4^2+e_5^2+e_6^2\right)\notag\\
&\phantom{:}+B_6(c)(e_{2,1}^2+e_{2,2}^2+e_{2,3}^2+e_{3,1}^2+e_{3,2}^2+e_{3,3}^2),
\label{E:Psi_e}
\end{align}
\label{E:Psi}%
\end{subequations}
where $A_1$ and $A_2$ are positive constants, $e_{\text{chem}}(c),B_1(c),...,B_7(c)$ are polynomial functions of $c$, among which $B_1$, $B_4$, $B_5$, and $B_6$ are positive, $K_{AB}(c)$ are also polynomial functions representing the components of the positive-definite, symmetric tensor that penalizes composition gradients, and finally $e_1,...,e_6$ are reparameterized strains defined as:
\begin{subequations}
\begin{align}
e_1&=(E_{11}+E_{22}+E_{33})/\sqrt{3},\label{E:ea}\\
e_2&=(E_{11}-E_{22})/\sqrt{2},\\
e_3&=(E_{11}+E_{22}-2E_{33})/\sqrt{6},\\
e_4&=E_{23}=E_{32},\\
e_5&=E_{13}=E_{31},\\
e_6&=E_{12}=E_{21},\label{E:ef}
\end{align}
\label{E:e}%
\end{subequations}
where $E_{IJ}=1/2(F_{kI}F_{kJ}-\delta_{IJ})$ are the components of the Green-Lagrange strain tensor.  
\begin{figure}
\centering\includegraphics[scale=1]{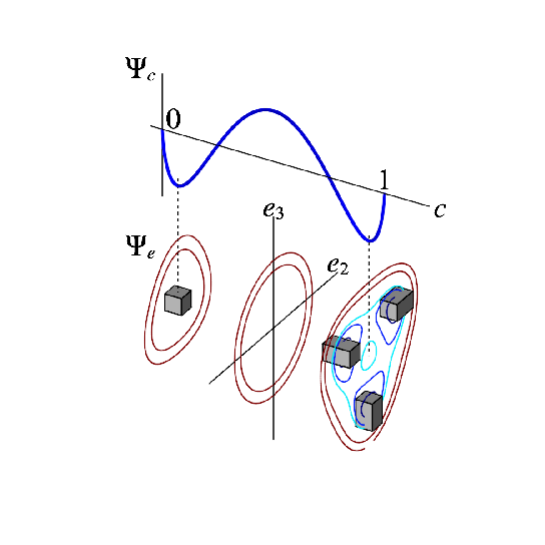}
\caption{Plots of $\Psi_c$ ($c\in(0,1)$) and contour plots of $\Psi_e$ on the $e_2-e_3$ plane at values of $c$ corresponding to the maximum and the two minima of $\Psi_c$ for a select set of parameters.  Stable cubic/tetragonal crystal structures are also depicted.  }
\label{Fi:plot_Psic_Psie_3D}
\end{figure}
Fig. \ref{Fi:plot_Psic_Psie_3D} shows, for a select set of parameters, plots of $\Psi_c$ ($c\in(0,1)$) and contour plots of $\Psi_e$ on the $e_2-e_3$ plane at values of $c$ corresponding to the maximum and the two minima of $\Psi_c$ along with stable cubic/tetragonal crystal structures.  
To facilitate formulation and analysis, we denote by $\bs{\zeta}$ an array of $c$, $c_{,A}$, $F_{iJ}$, and $F_{iJ,K}$ and define a new multivariate scalar function $\Psi_{s+e}:=\Psi_s+\Psi_e$.  
Note that, from definitions \eqref{E:Psi_s}, \eqref{E:Psi_e}, and \eqref{E:e}, $\Psi_{s+e}$ is a multivariate polynomial function of $c$, $c_{,A}$, $F_{iJ}$, and $F_{iJ,K}$, which is crucial for developing our accurate time-integration algorithm.  

    \subsection{Chemical equilibrium/non-equilibrium}\label{SS:variational_formulation_chemistry}
    In this section we derive the equations for non-equilibrium chemistry.  
We first take the variational derivative of the total free-energy functional \eqref{E:Pi} with respect to the chemical composition $c$ in the direction of $q$ to obtain:
\begin{align}
\delta_c\Pi[c,\bs{u}]
=\!\left.\frac{\dif}{\dif\varepsilon}\Pi[c+\varepsilon q,\bs{u}]\right|_{\varepsilon=0}
=\int_{\Omega}
q
(\bar{\mu}(c)+H(\bs{\zeta}))
+q_{,A}W_A(\bs{\zeta})
\dif V,\label{E:c_equil_weak_temp}
\end{align}
where $\bar{\mu}(c)$, $H(\bs{\zeta})$, and $W_A(\bs{\zeta})$ are defined as:
\begin{align}
\bar{\mu}&:=\frac{\dif\Psi_c}{\dif c},\label{E:mubar}\\
H&:=\frac{\partial\Psi_{s+e}}{\partial c},\label{E:H}\\
W_A&:=\frac{\partial\Psi_{s+e}}{\partial c_{,A}},\label{E:W_A}
\end{align}
where $\bar{\mu}(c)$ is known as the homogeneous chemical potential.  
At equilibrium one has $\delta_c\Pi[c,\bs{u}]=0$ and therefore from \eqref{E:c_equil_weak_temp} one obtains the chemical-equilibrium equation as:
\begin{align}
\int_{\Omega}q
\left(
\bar{\mu}(c)+H(\bs{\zeta})
\right)
+q_{,A}W_A(\bs{\zeta})
\dif V
=0.\label{E:c_equil_weak}
\end{align}
We then apply the divergence theorem to Eqn. \eqref{E:c_equil_weak} to obtain:
\begin{align}
\int_{\Omega}q
\left(
\bar{\mu}+H-{W_A}_{,A}
\right)
\dif V
+\int_{\Gamma}qW_AN_A\dif S
=0.\notag
\end{align}
Standard variational arguments then lead us to the following strong form for chemical equilibrium and the corresponding boundary condition:
\begin{subequations}
\begin{align}
\bar{\mu}+H-{W_A}_{,A}&=0\quad\text{in }\Omega,\label{E:c_equil_strong_vol}\\
W_AN_A&=0\quad\on\Gamma\label{E:c_equil_strong_bc}.
\end{align}
\label{E:c_equil_strong}%
\end{subequations}
Note that one can further write Eqns. \eqref{E:c_equil_strong_vol} and \eqref{E:c_equil_strong_bc} as:
\begin{align}
\bar{\mu}+H-(K_{AB}c_{,B})_{,A}&=0\quad\text{in }\Omega,\notag\\
(K_{AB}c_{,B})N_A&=0\quad\on\Gamma,\notag
\end{align}
but in this work we avoid this step to simplify our stability analysis presented in Sec.\ref{SS:analysis_stability}.  
We identify the left-hand side of Eqn. \eqref{E:c_equil_strong_vol} as the \emph{chemical potential} $\mu$, i.e.:
\begin{align}
\mu=\bar{\mu}+H-{W_A}_{,A}.\label{E:mu}
\end{align}
With the expression for the chemical potential \eqref{E:mu} in hand, one can formulate the non-equilibrium chemistry problem using the mass balance law in conjunction with the phenomenological representation of the flux, that is:
\begin{align}
&\frac{\D c}{\D t}+j_{A,A}=0,\label{E:c_nonequil_strong}
\end{align}
where, as elsewhere,  $\D/\D t$ represents the material time-derivative, and the flux is in coordinate notation,
\begin{align}
&j_A=-L_{AB}\mu_{,B},\label{flux_phenomenological}
\end{align}
with $L_{AB}(c)$ being the components of the positive definite mobility tensor.  
Eqn.\eqref{E:c_nonequil_strong} requires another boundary condition for $j_AN_A$ on $\Gamma$.  
Throughout this work we follow the proposition in \cite{Gomez2011} and set:
\begin{align}
j_AN_A=0\quad\on\Gamma.\label{E:c_nonequil_strong_bc}
\end{align}
In this work we adopt the mixed formulation in which $\mu$ as well as $c$ are regarded as primary unknowns.  
Multiplying Eqns. \eqref{E:c_nonequil_strong} and \eqref{E:mu} by admissible test functions $q$ and $\nu$, applying the divergence theorem applying boundary conditions \eqref{E:c_equil_strong_bc} and \eqref{E:c_nonequil_strong_bc}, one obtains the weak form for the non-equilibrium chemistry problem as:
\begin{subequations}
\begin{align}
&\int_{\Omega}\left(q\frac{\D c}{\D t}+q_{,A}L_{AB}(c)\mu_{,B}\right)
\dif V
=0,\label{E:c_nonequil_weak}\\
&\int_{\Omega}
\left(\nu\left(
-\mu
+\bar{\mu}(c)
+H(\bs{\zeta})\right)
+\nu_{,A}W_A(\bs{\zeta})\right)
\dif V
=0.\label{E:mu_nonequil_weak}
\end{align}
\label{E:cmu_nonequil_weak}%
\end{subequations}

    \subsection{Mechanical equilibrium}\label{SS:variational_formulation_mechanics}
    To formulate the problem of mechanical equilibrium, we take the variational derivative of the total free energy \eqref{E:Pi} with respect to $\bs{u}$ that satisfies the Dirichlet boundary conditions \eqref{E:u_DirichletBC}.  
The test function $\bs{w}$ is then to satisfy:
\begin{align}
w_i=0\quad\on\Gamma^u,\quad
Dw_i=0\quad\on\Gamma^m,\quad
w_i=0\quad\on\Upsilon^g.\notag
\end{align}
The variational derivative with respect to $\bs{u}$ is then obtained as:
\begin{align}
\delta_{\bs{u}}\Pi[c,\bs{u}]
&=\!\left.\frac{\dif}{\dif\varepsilon}\Pi[c,\bs{u}+\varepsilon\bs{w}]\right|_{\varepsilon=0}\notag\\
&=\int_{\Omega}\left(w_{i,J}P_{iJ}(\bs{\zeta})+w_{i,JK}B_{iJK}(\bs{\zeta})\right)\dif V
-\int_{\Gamma^T}w_i\bar{T}_i\dif S
-\int_{\Gamma^M}Dw_i\bar{M}_i\dif S
-\int_{\Upsilon^G}w_i\bar{G}_i\dif C,\label{E:u_equil_weak_temp}
\end{align}
where $P_{iJ}(\bs{\zeta})$ are the components of the first Piola-Kirchhoff stress tensor and $B_{iJK}(\bs{\zeta})$ are the components of the higher-order stress tensor that are defined as:
\begin{align}
P_{iJ}&:=\frac{\partial\Psi_{s+e}}{\partial F_{iJ}},\notag\\
B_{iJK}&:=\frac{\partial\Psi_{s+e}}{\partial F_{iJ,K}}.\notag
\end{align}
At equilibrium one has $\delta_{\bs{u}}\Pi[c,\bs{u}]=0$ so that from \eqref{E:u_equil_weak_temp} one obtains:
\begin{align}
\int_{\Omega}\left(w_{i,J}P_{iJ}(\bs{\zeta})+w_{i,JK}B_{iJK}(\bs{\zeta})\right)\dif V
-\int_{\Gamma^T}w_i\bar{T}_i\dif S
-\int_{\Gamma^M}Dw_i\bar{M}_i\dif S
-\int_{\Upsilon^G}w_i\bar{G}_i\dif C=0.\label{E:u_equil_weak}
\end{align}
Eqn. \eqref{E:u_equil_weak} is the weak form for mechanical equilibrium that is to be solved in conjunction with the weak form for the non-equilibrium problem of chemistry \eqref{E:cmu_nonequil_weak}.  

The variational argument can further lead us to identify the strong form and the Neumann boundary conditions corresponding to \eqref{E:u_equil_weak} as the following:
\begin{align}
P_{iJ,J}-B_{iJK,JK}&=0\hfill                                                                              &&\text{in }\Omega,\notag\\
P_{iJ}N_J-B_{iJK,K}N_J-D_J(B_{iJK}N_K)+B_{iJK}\left(b_{LL} N_JN_K-b_{JK}\right)&=\bar{T}_i   &&\on\Gamma^T,\notag\\
B_{iJK}N_KN_J&=\bar{M}_i                                                                                  &&\on\Gamma^M,\notag\\
[\![B_{iJK}N_KN_J^\Gamma]\!]&=\bar{G}_i                                        &&\on\Upsilon^G,\notag
\end{align}
where $b_{IJ}$ are the components of the second fundamental form on $\Gamma^T$, $N^{\Gamma}_J$ are the components of the unit outward normal to the boundary curve $\Upsilon_{\iota}\subset\overline{\Gamma_{\iota'}}$, and, on each $\Upsilon^G_\iota$, $[\![B_{iJK}N_KN_J^\Gamma]\!]:=B_{iJK}N^+_KN_J^{\Gamma+}+B_{iJK}N^-_KN_J^{\Gamma-}$ is the \emph{jump}, where superscripts $+$ and $-$ represent two surfaces sharing $\Upsilon^G_\iota$; see \cite{Toupin1964} for details.

\section{Numerical formulations}\label{S:numerical_formulation}
    \subsection{Spatial discretization}\label{SS:numerical_formulation_space}
    We now discretize Eqns. \eqref{E:cmu_nonequil_weak} and \eqref{E:u_equil_weak} in space for formulations that are amenable to numerical analysis.  
Since $\Psi_{s+e}$ is a function of $F_{iJ,K}$, our weak forms \eqref{E:cmu_nonequil_weak} and \eqref{E:u_equil_weak} involve second-order spatial derivatives of the displacement field $\bs{u}$ and the test function $\bs{w}$, and thus require them to be $\mathcal{W}^{2,2}$, where $\mathcal{W}^{s,p}$ is the standard Sobolev space.   
We denote by $\mathcal{S}^h$ an appropriate finite-dimensional subspace of $\mathcal{W}^{2,2}(\ol{\Omega})$ and define:  
\begin{align}
\mathcal{V}_u^h&=\left\{\bs{u}^h\in[\mathcal{S}^h]^3:u_i^h=\bar{u}_i\quad\on\Gamma^u,\quad Du_i^h=\bar{m}_i\quad\on\Gamma^m,\quad u_i^h=\bar{g}_i\quad\on\Upsilon^g\right\},\notag\\
\mathcal{V}_w^h&=\left\{\bs{w}^h\in[\mathcal{S}^h]^3:w_i^h=0\quad\on\Gamma^u,\quad Dw_i^h=0\quad\on\Gamma^m,\quad w_i^h=0\quad\on\Upsilon^g\right\},\notag
\end{align}
assuming that $\mathcal{S}^h$ allows for exact representation of the Dirichlet boundary conditions \eqref{E:u_DirichletBC}.  
We also denote by $\mathcal{T}^h$ an appropriate finite-dimensional subspace of $\mathcal{W}^{1,2}(\ol{\Omega})$.  

The space-discrete counterparts of the weak formulations \eqref{E:cmu_nonequil_weak} and \eqref{E:u_equil_weak} are then given as the following:

$  $\\
Seek $c^h(\bs{X},t)\in\mathcal{T}^h\times[0,T]$, $\mu^h(\bs{X},t)\in\mathcal{T}^h\times[0,T]$, $\bs{u}^h(\bs{X},t)\in\mathcal{V}_u^h\times[0,T]$ such that for all $q^h(\bs{X})\in\mathcal{T}^h$, $\nu^h(\bs{X})\in\mathcal{T}^h$, $\bs{w}^h(\bs{X})\in\mathcal{V}_w^h$:
\begin{subequations}
\begin{align}
&\int_{\Omega}\left(q^h\frac{Dc^h}{Dt}+q^h_{,A}L_{AB}(c^h)\mu^h_{,B}\right)\dif V=0,\label{E:c_space}\\
&\int_{\Omega}\left(\nu^h\left(-\mu^h+\bar{\mu}(c^h)+H(\bs{\zeta}^h)\right)+\nu_{,A}^hW_A(\bs{\zeta}^h)\right)\dif V=0,\label{E:mu_space}\\
&\int_{\Omega}\left(w_{i,J}^hP_{iJ}(\bs{\zeta}^h)+w_{i,JK}^hB_{iJK}(\bs{\zeta}^h)\right)\dif V
-\int_{\Gamma^T}w^h_i\bar{T}_i\dif S
-\int_{\Gamma^M}Dw^h_i\bar{M}_i\dif S
-\int_{\Upsilon^G}w^h_i\bar{G}_i\dif C
=0,\label{E:u_space}
\end{align}
\label{E:cmuu_space}%
\end{subequations}
where $\bs{\zeta}^h(\bs{X},t)$ is an array of $c^h$, $c^h_{,A}$, $F^h_{iJ}(=\delta_{iJ}+u^h_{i,J})$, and $F^h_{iJ,K}$. Here, all spatial derivatives are now to be understood in the weak sense.\footnote{Since $\mathcal{S}^h\subset\mathcal{W}^{2,p}(\ol{\Omega})$, second spatial derivatives of $\bs{w}^h$ and $\bs{u}^h$ are properly defined only in a weak sense---a technicality that is usually not emphasized in finite dimensional weak forms.}

Note that, setting $q^h=1$, $\nu^h=0$, and $\bs{w}^h=\bs{0}$ in the weak form \eqref{E:cmuu_space}, one recovers the mass conservation law:
\begin{align}
\frac{\dif}{\dif t}\int_\Omega c^h\dif V=0.\notag
\end{align}
On the other hand, assuming that all Dirichlet and Neumann boundary conditions are time-independent and setting $q^h=\mu^h$, $\nu^h=\D c^h/\D t$, and $\bs{w}^h=\D \bs{u}^h/\D t$ in Eqns. \eqref{E:cmuu_space} and adding them together, one obtains:
\begin{align}
\frac{\dif\Pi^h}{\dif t}=-\int_\Omega \mu^h_{,A}L_{AB}\mu^h_{,B}\dif V,\label{E:dPidt}
\end{align}
where $\Pi^h$ is the space-discrete total free energy at arbitrary time $t$ defined as:
\begin{align}
\Pi^{h}(t)
&=\int_{\Omega}\left(
\Psi_{c}(c^h)
+\Psi_{s+e}(\bs{\zeta}^h)\right)\dif V
-\int_{\Gamma^T}u^h_i\bar{T}_i\dif S
-\int_{\Gamma^M}Du^h_i\bar{M}_i\dif S
-\int_{\Upsilon^G}u^h_i\bar{G}_i\dif C,\label{E:Pi^h}
\end{align}
spatial derivatives being understood in the weak sense.  
Recalling that the mobility tensor is positive definite, Eqn.\eqref{E:dPidt} implies non-increasing free energy.  

These two properties, mass conservation and non-increasing total free energy, are to be inherited by our space-time discrete formulation developed in Sec. \ref{SS:numerical_formulation_time}; 
specifically, the latter property furnishes the notion of \emph{stability}.

As is well known, substituting Eqns. (\ref{flux_phenomenological}), (\ref{E:mu}) and (\ref{E:mubar}--\ref{E:W_A}) into (\ref{E:c_nonequil_strong}) leads to a fourth-order PDE in strong form. Its weak counterpart has up to second-order spatial derivatives on the composition, and was the basis for Discontinuous Galerkin-based finite element methods in the work of Wells et al. \cite{Wellsetal2006}, and more recently for $C^1$-continuous IGA-based methods in \cite{Rudrarajuetal2015}. 
Here, we use the split formulation of Eqn. (\ref{E:cmu_nonequil_weak}), because, as shown above, it permits the fields $c^h$ and $\mu^h$ to be chosen to lie in the same space, $\mathcal{T}^h$, in the resulting finite dimensional statement of the full problem (\ref{E:cmuu_space}). 
This coincidence of spaces is crucial for satisfaction of the fundamental stability result just derived, and for its extension to the time-discrete setting.

    \subsection{Temporal discretization}\label{SS:numerical_formulation_time}
    We proceed to discretize Eqns. \eqref{E:cmuu_space} in time to obtain a formulation that produces a solution at time $t^{n+1}$ given a solution at time $t^n$.  
The proposed time-discrete formulation is given as the following:

$  $\\
Given $c^{h,n}(\bs{X})\in\mathcal{T}^h$, $\mu^{h,n}(\bs{X})\in\mathcal{T}^h$, $\bs{u}^{h,n}(\bs{X})\in\mathcal{V}_u^h$, seek $c^{h,n+1}(\bs{X})\in\mathcal{T}^h$, $\mu^{h,n+1}(\bs{X})\in\mathcal{T}^h$, $\bs{u}^{h,n+1}(\bs{X})\in\mathcal{V}_u^h$ such that for all $q^h(\bs{X})\in\mathcal{T}^h$, $\nu^h(\bs{X})\in\mathcal{T}^h$, $\bs{w}^h(\bs{X})\in\mathcal{V}_w^h$:
\begin{subequations}
\begin{align}
&\int_{\Omega}\left(q^h\left\{\!\frac{Dc^h}{Dt}\!\right\}^n+q^h_{,A}\{L_{AB}(c^h)\}^n\{\mu^h\}^n_{,B}\right)\dif V=0,\label{E:c_space_time}\\
&\int_{\Omega}\left(\nu^h\left(-\{\mu^h\}^n+\{\bar{\mu}(c^h)\}^n+\{H(\bs{\zeta}^h)\}^n\right)+\nu_{,A}^h\{W_A(\bs{\zeta}^h)\}^n\right)\dif V=0,\label{E:mu_space_time}\\
&\int_{\Omega}\left(w_{i,J}^h\{P_{iJ}(\bs{\zeta}^h)\}^n+w_{i,JK}^h\{B_{iJK}(\bs{\zeta}^h)\}^n\right)\dif V\notag\\
&\hspace{100pt}
-\int_{\Gamma^T}w^h_i\left\{\bar{T}_i\right\}^n\dif S
-\int_{\Gamma^M}Dw^h_i\left\{\bar{M}_i\right\}^n\dif S
-\int_{\Upsilon^G}w^h_i\left\{\bar{G}_i\right\}^n\dif C=0,\label{E:u_space_time}
\end{align}
\label{E:cmuu_space_time}%
\end{subequations}
where terms with braces $\{\hspace{1pt}\bullet\hspace{1pt}\}^n(\bs{X})$ represent time-discretizations of those quantities inside the braces $\bullet\hspace{2pt}(\bs{X},t)$ on the time-interval $t\in[t^n,t^{n+1}]$, and they are defined in the rest of this section so that the formulation in \eqref{E:cmuu_space_time} is second-order accurate and unconditionally stable.  

The stability analysis presented in Sec. \ref{SS:analysis_stability} is greatly motivated by that presented in \cite{Gomez2011} for the Cahn-Hilliard equation, where $\{\bar{\mu}(c^h)\}^n$ was defined using dedicated quadrature formulas so that a non-increasing chemical free energy would be the direct consequence of the weak formulation.  
In this work we follow the same course, but, instead of developing special quadrature formulas, we employ Taylor expansions, which enables one to simplify the argument as well as to extend the stability analysis to coupling with gradient elasticity.  

For convenience we denote by $\bs{\zeta}^{h,n}(\bs{X})$ the temporal approximation to $\bs{\zeta}^h(\bs{X},t^n)$.  
In addition we define $\Delta t:=t^{n+1}-t^{n}$, $\Delta c^h:=c^{h,n+1}-c^{h,n}$, $\Delta c^h_{,A}:=c_{,A}^{h,n+1}-c_{,A}^{h,n}$, $\Delta F^h_{iJ}:=u_{i,J}^{h,n+1}-u_{i,J}^{h,n}$, and $\Delta F^h_{iJ,K}:=u_{i,JK}^{h,n+1}-u_{i,JK}^{h,n}$.  

We first represent $\{\bar{\mu}(c^h)\}^n$ in terms of $c^{h,n}$ and $c^{h,n+1}$ at each fixed point $\bs{X}\in\Omega$.  
To this end we observe that the Taylor expansion of $\Psi_c(c)$ around $c^{h,n+1}$ leads to the following identity:
\begin{align}
\Psi_c(c^{h,n})
&=\Psi_c(c^{h,n+1})
-\bar{\mu}(c^{h,n+1})\Delta c^h
+\frac{1}{2}\frac{\dif\bar{\mu}}{\dif c}(c^{h,n+1})(\Delta c^h)^2
-\frac{1}{6}\frac{\dif^2\bar{\mu}}{\dif c^2}(c^{h,n+1})(\Delta c^h)^3
+\frac{1}{24}\frac{\dif^3\bar{\mu}}{\dif c^3}(\xi)(\Delta c^h)^4\notag\\
&=\Psi_c(c^{h,n+1})
-\left(
\bar{\mu}(c^{h,n+1})
-\frac{1}{2}\frac{\dif\bar{\mu}}{\dif c}(c^{h,n+1})\Delta c^h
+\frac{1}{6}\frac{\dif^2\bar{\mu}}{\dif c^2}(c^{h,n+1})(\Delta c^h)^2
\right)\Delta c^h
+\frac{1}{24}\frac{\dif^3\bar{\mu}}{\dif c^3}(\xi)(\Delta c^h)^4,
\label{E:Psic_Taylor}
\end{align}
where $\xi=(1-\alpha)c^{h,n}+\alpha c^{h,n+1}$ for some $\alpha$ ($0<\alpha<1$) by Taylor's Remainder Theorem.  
We then define $\{\bar{\mu}(c^h)\}^n$ as the quantity in the parentheses in \eqref{E:Psic_Taylor}, that is:
\begin{align}
\{\bar{\mu}(c^h)\}^n
:=\bar{\mu}(c^{h,n+1})
-\frac{1}{2}\frac{\dif\bar{\mu}}{\dif c}(c^{h,n+1})\Delta c^h
+\frac{1}{6}\frac{\dif^2\bar{\mu}}{\dif c^2}(c^{h,n+1})(\Delta c^h)^2,
\label{E:{mubar}}
\end{align}
so that:
\begin{align}
\{\bar{\mu}(c^h)\}^n\Delta c^h=
\Psi_c(c^{h,n+1})-\Psi_c(c^{h,n})
+\frac{1}{24}\frac{\dif^3\bar{\mu}}{\dif c^3}\left(\xi\right)(\Delta c^h)^4,
\label{E:{mubar}deltac}
\end{align}
at each fixed point $\bs{X}\in\Omega$.  
Identity \eqref{E:{mubar}deltac} becomes a convenient tool in the stability analysis encountered in Sec. \ref{S:analysis}, noting especially that $\dif^3\bar{\mu}/\!\!\dif c^3(\xi)>0$ by virtue of \eqref{E:Psi_c}.

We define $\{H(\bs{\zeta}^h)\}^n$, $\{W_A(\bs{\zeta}^h)\}^n$, $\{P_{iJ}(\bs{\zeta}^h)\}^n$, and $\{B_{iJK}(\bs{\zeta}^h)\}^n$ in a similar fashion.  
We first denote by $\mathcal{D}\left[\phi;\kappa_c,\kappa_{\nabla c},\kappa_F,\kappa_{\nabla F}\right]$ the function obtained by applying operators $(\partial/\partial c)\Delta c^h$, $(\partial/\partial c_{,A})\Delta c^h_{,A}$, $(\partial/\partial F_{iJ})\Delta F^h_{iJ}$, and $(\partial/\partial F_{iJ,K})\Delta F^h_{iJ,K}$ respectively $\kappa_c$, $\kappa_{\nabla c}$, $\kappa_F$, and $\kappa_{\nabla F}$ ($\kappa_c$, $\kappa_{\nabla c}$, $\kappa_F$, $\kappa_{\nabla F}\geq0$) times to a scalar-valued multivariate function $\phi(\bs{\zeta})$.  
For instance we have:
\begin{align*}
\mathcal{D}\left[\phi;0,0,0,0\right]&=\phi,\\
\mathcal{D}\left[\phi;1,0,2,1\right]&=\frac{\partial^4\phi^n}{\partial c\partial F_{iJ}\partial F_{kL}\partial F_{mN,O}}\Delta c^h\Delta F^h_{iJ}\Delta F^h_{kL}\Delta F^h_{mN,O}
=\mathcal{D}\left[\frac{\partial\phi}{\partial F_{iJ}};1,0,1,1\right]\Delta F^h_{iJ}.
\end{align*}
We also define $\kappa=\kappa_c+\kappa_{\nabla c}+\kappa_F+\kappa_{\nabla F}$.  
The Taylor expansion of $\Psi_{s+e}(\bs{\zeta})$ around $\bs{\zeta}^{h,n}(\bs{X})$ at a fixed point $\bs{X}\in\Omega$ then leads to the following identity:
\begin{align}
\Psi_{s+e}(\bs{\zeta}^{h,n+1})
=\Psi_{s+e}(\bs{\zeta}^{h,n})
&+\sum_{\kappa\geq 1}\frac{\kappa !}{\kappa_c!\kappa_{\nabla c}!\kappa_F!\kappa_{\nabla F}!}\frac{1}{\kappa !}\mathcal{D}\left[\Psi_{s+e};\kappa_c,\kappa_{\nabla c},\kappa_F,\kappa_{\nabla F}\right](\bs{\zeta}^{h,n})\notag\\
=\Psi_{s+e}(\bs{\zeta}^{h,n})
&+\sum_{\substack{\kappa_c\geq 1\\\kappa\ge 1}}\frac{\kappa_c}{\kappa}\frac{1}{\kappa_c!\kappa_{\nabla c}!\kappa_F!\kappa_{\nabla F}!}\mathcal{D}\left[\Psi_{s+e};\kappa_c,\kappa_{\nabla c},\kappa_F,\kappa_{\nabla F}\right](\bs{\zeta}^{h,n})\notag\\
&+\sum_{\substack{\kappa_{\nabla c}\geq 1\\\kappa\ge 1}}\frac{\kappa_{\nabla c}}{\kappa}\frac{1}{\kappa_c!\kappa_{\nabla c}!\kappa_F!\kappa_{\nabla F}!}\mathcal{D}\left[\Psi_{s+e};\kappa_c,\kappa_{\nabla c},\kappa_F,\kappa_{\nabla F}\right](\bs{\zeta}^{h,n})\notag\\
&+\sum_{\substack{\kappa_F\geq 1\\\kappa\ge 1}}\frac{\kappa_F}{\kappa}\frac{1}{\kappa_c!\kappa_{\nabla c}!\kappa_F!\kappa_{\nabla F}!}\mathcal{D}\left[\Psi_{s+e};\kappa_c,\kappa_{\nabla c},\kappa_F,\kappa_{\nabla F}\right](\bs{\zeta}^{h,n})\notag\\
&+\sum_{\substack{\kappa_{\nabla F}\geq 1\\\kappa\ge 1}}\frac{\kappa_{\nabla F}}{\kappa}\frac{1}{\kappa_c!\kappa_{\nabla c}!\kappa_F!\kappa_{\nabla F}!}\mathcal{D}\left[\Psi_{s+e};\kappa_c,\kappa_{\nabla c},\kappa_F,\kappa_{\nabla F}\right](\bs{\zeta}^{h,n})\notag\\
=\Psi_{s+e}(\bs{\zeta}^{h,n})
&+\left(\sum_{\substack{\kappa_c\geq 1\\\kappa\ge 1}}\frac{1}{\kappa}\frac{1}{\left(\kappa_c-1\right)!\kappa_{\nabla c}!\kappa_F!\kappa_{\nabla F}!}\mathcal{D}\left[H;\kappa_c-1,\kappa_{\nabla c},\kappa_F,\kappa_{\nabla F}\right](\bs{\zeta}^{h,n})\right)\Delta c^h\notag\\
&+\left(\sum_{\substack{\kappa_{\nabla c}\geq 1\\\kappa\ge 1}}\frac{1}{\kappa}\frac{1}{\kappa_c!\left(\kappa_{\nabla c}-1\right)!\kappa_F!\kappa_{\nabla F}!}\mathcal{D}\left[W_A;\kappa_c,\kappa_{\nabla c}-1,\kappa_F,\kappa_{\nabla F}\right](\bs{\zeta}^{h,n})\right)\Delta c^h_{,A}\notag\\
&+\left(\sum_{\substack{\kappa_F\geq 1\\\kappa\ge 1}}\frac{1}{\kappa}\frac{1}{\kappa_c!\kappa_{\nabla c}!\left(\kappa_F-1\right)!\kappa_{\nabla F}!}\mathcal{D}\left[P_{iJ};\kappa_c,\kappa_{\nabla c},\kappa_F-1,\kappa_{\nabla F}\right](\bs{\zeta}^{h,n})\right)\Delta F^h_{iJ}\notag\\
&+\left(\sum_{\substack{\kappa_{\nabla F}\geq 1\\\kappa\ge 1}}\frac{1}{\kappa}\frac{1}{\kappa_c!\kappa_{\nabla c}!\kappa_F!\left(\kappa_{\nabla F}-1\right)!}\mathcal{D}\left[B_{iJK};\kappa_c,\kappa_{\nabla c},\kappa_F,\kappa_{\nabla F}-1\right](\bs{\zeta}^{h,n})\right)\Delta F^h_{iJ,K},
\label{E:Psis+e_Taylor}
\end{align}
where summations are over all possible combinations of $\kappa_c$, $\kappa_{\nabla c}$, $\kappa_F$, and $\kappa_{\nabla F}$ for each $\kappa$.  
These summations are finite as $\Psi_{s+e}(\bs{\zeta})$ is a multivariate polynomial function of $c$, $c_{,A}$, $F_{iJ}$, and $F_{iJ,K}$.  
Factors in the summation in the first line arise since $\mathcal{D}\left[\Psi_{s+e};\kappa_c,\kappa_{\nabla c},\kappa_F,\kappa_{\nabla F}\right](\bs{\zeta}^{h,n})$ appears in a straightforward Taylor-series expansion $\kappa !/\kappa_{c}!\kappa_{\nabla c}!\kappa_{F}!\kappa_{\nabla F}!$ times due to this number of possible permutations; for the sufficiently smooth , $\Psi$ considered here, the following terms all reduce to $(1/3!)\cdot\mathcal{D}\left[\Psi_{s+e};0,0,2,1\right](\bs{\zeta}^{h,n})$ and therefore this term in the above summation is to be multiplied by $3!/0!0!2!1!=3$:
\begin{align*}
    &\frac{1}{3!}\frac{\partial^3\Psi_{s+e}}{\partial F_{iJ}\partial F_{kL}\partial F_{mN,O}}(\bs{\zeta}^{h,n})\Delta F_{iJ}\Delta F_{kL}\Delta F_{mN,O},\\
    &\frac{1}{3!}\frac{\partial^3\Psi_{s+e}}{\partial F_{iJ}\partial F_{mN,O}\partial F_{kL}}(\bs{\zeta}^{h,n})\Delta F_{iJ}\Delta F_{mN,O}\Delta F_{kL},\\
    &\frac{1}{3!}\frac{\partial^3\Psi_{s+e}}{\partial F_{mN,O}\partial F_{iJ}\partial F_{kL}}(\bs{\zeta}^{h,n})\Delta F_{mN,O}\Delta F_{iJ}\Delta F_{kL}.
\end{align*}
We then define $\{H(\bs{\zeta}^h)\}^n$, $\{W_A(\bs{\zeta}^h)\}^n$, $\{P_{iJ}(\bs{\zeta}^h)\}^n$, and $\{B_{iJK}(\bs{\zeta}^h)\}^n$ as those quantities in the parentheses in \eqref{E:Psis+e_Taylor}, or:
\begin{subequations}
\begin{align}
&\{H(\bs{\zeta}^h)\}^n\notag\\
&:=H(\bs{\zeta}^{h,n})
+\frac{1}{2}\left(
\frac{\partial H}{\partial c}(\bs{\zeta}^{h,n})\Delta c^h
+\frac{\partial H}{\partial c_{,B}}(\bs{\zeta}^{h,n})\Delta c^h_{,B}
+\frac{\partial H}{\partial F_{lM}}(\bs{\zeta}^{h,n})\Delta F^h_{lM}
+\frac{\partial H}{\partial F_{lM,N}}(\bs{\zeta}^{h,n})\Delta F^h_{lM,N}
\right)
+R^{c}(\bs{\zeta}^{h,n}),\label{E:{H}}\\
&\{W_A(\bs{\zeta}^h)\}^n\notag\\
&:=W_A(\bs{\zeta}^{h,n})
+\frac{1}{2}\left(
\frac{\partial W_A}{\partial c}(\bs{\zeta}^{h,n})\Delta c^h
+\frac{\partial W_A}{\partial c_{,B}}(\bs{\zeta}^{h,n})\Delta c^h_{,B}
+\frac{\partial W_A}{\partial F_{lM}}(\bs{\zeta}^{h,n})\Delta F^h_{lM}
+\frac{\partial W_A}{\partial F_{lM,N}}(\bs{\zeta}^{h,n})\Delta F^h_{lM,N}
\right)
+R^{\nabla c}_A(\bs{\zeta}^{h,n}),\label{E:{W}}\\
&\{P_{iJ}(\bs{\zeta}^h)\}^n\notag\\
&:=P_{iJ}(\bs{\zeta}^{h,n})
+\frac{1}{2}\left(
\frac{\partial P_{iJ}}{\partial c}(\bs{\zeta}^{h,n})\Delta c^h
+\frac{\partial P_{iJ}}{\partial c_{,B}}(\bs{\zeta}^{h,n})\Delta c^h_{,B}
+\frac{\partial P_{iJ}}{\partial F_{lM}}(\bs{\zeta}^{h,n})\Delta F^h_{lM}
+\frac{\partial P_{iJ}}{\partial F_{lM,N}}(\bs{\zeta}^{h,n})\Delta F^h_{lM,N}
\right)
+R^{F}_{iJ}(\bs{\zeta}^{h,n}),\label{E:{P}}\\
&\{B_{iJK}(\bs{\zeta}^h)\}^n\notag\\
&:=B_{iJK}(\bs{\zeta}^{h,n})
+\frac{1}{2}\left(
\frac{\partial B_{iJK}}{\partial c}(\bs{\zeta}^{h,n})\Delta c^h
+\frac{\partial B_{iJK}}{\partial c_{,B}}(\bs{\zeta}^{h,n})\Delta c^h_{,B}
+\frac{\partial B_{iJK}}{\partial F_{lM}}(\bs{\zeta}^{h,n})\Delta F^h_{lM}
+\frac{\partial B_{iJK}}{\partial F_{lM,N}}(\bs{\zeta}^{h,n})\Delta F^h_{lM,N}
\right)
+R^{\nabla F}_{iJK}(\bs{\zeta}^{h,n}),\label{E:{B}}
\end{align}
\label{E:{HWPB}}%
\end{subequations}
where
\begin{subequations}
\begin{align}
R^{c}:&=\sum_{\substack{\kappa_c\geq 1\\\kappa\geq3}}\frac{1}{\kappa}\frac{1}{\left(\kappa_c-1\right)!\kappa_{\nabla c}!\kappa_F!\kappa_{\nabla F}!}\mathcal{D}\left[H;\kappa_c-1,\kappa_{\nabla c},\kappa_F,\kappa_{\nabla F}\right],\\
R^{\nabla c}_A:&=\sum_{\substack{\kappa_{\nabla c}\geq 1\\\kappa\geq3}}\frac{1}{\kappa}\frac{1}{\kappa_c!\left(\kappa_{\nabla c}-1\right)!\kappa_F!\kappa_{\nabla F}!}\mathcal{D}\left[W_A;\kappa_c,\kappa_{\nabla c}-1,\kappa_F,\kappa_{\nabla F}\right],\\
R^{F}_{iJ}:&=\sum_{\substack{\kappa_F\geq 1\\\kappa\geq3}}\frac{1}{\kappa}\frac{1}{\kappa_c!\kappa_{\nabla c}!\left(\kappa_F-1\right)!\kappa_{\nabla F}!}\mathcal{D}\left[P_{iJ};\kappa_c,\kappa_{\nabla c},\kappa_F-1,\kappa_{\nabla F}\right],\\
R^{\nabla F}_{iJK}:&=\sum_{\substack{\kappa_{\nabla F}\geq 1\\\kappa\geq3}}\frac{1}{\kappa}\frac{1}{\kappa_c!\kappa_{\nabla c}!\kappa_F!\left(\kappa_{\nabla F}-1\right)!}\mathcal{D}\left[B_{iJK};\kappa_c,\kappa_{\nabla c},\kappa_F,\kappa_{\nabla F}-1\right],
\end{align}
\label{E:Rs}%
\end{subequations}
so that:
\begin{align}
\{H(\bs{\zeta}^h)\}^n\Delta c^h
+\{W_A(\bs{\zeta}^h)\}^n\Delta c^h_{,A}
+\{P_{iJ}(\bs{\zeta}^h)\}^n\Delta F^h_{iJ}
+\{B_{iJK}(\bs{\zeta}^h)\}^n\Delta F^h_{iJ,K}
=\Psi_{s+e}(\bs{\zeta}^{h,n+1})-\Psi_{s+e}(\bs{\zeta}^{h,n}),
\label{E:{HWPB}delta}
\end{align}
at each fixed point $\bs{X}\in\Omega$.  
Finally, other quantities in Eqns.\eqref{E:cmuu_space_time} are defined as:
\begin{subequations}
\begin{align}
\left\{\!\frac{Dc^h}{Dt}\!\right\}^n&:=\frac{\Delta c^h}{\Delta t},\\
\{c^h\}^n&:=[c^{h,n+1}+c^{h,n}]/2,\\
\{\mu^h\}^n&:=[\mu^{h,n+1}+\mu^{h,n}]/2,\\
\{{L}_{AB}(c^h)\}^n&:=[{L}_{AB}(c^{h,n+1})+{L}_{AB}(c^{h,n})]/2,\\
\{\bar{T}_i\}^n&:=[\bar{T}_i^{n+1}+\bar{T}_i^n]/2,\\
\{\bar{M}_i\}^n&:=[\bar{M}_i^{n+1}+\bar{M}_i^n]/2,\\
\{\bar{G}_i\}^n&:=[\bar{G}_i^{n+1}+\bar{G}_i^n]/2,
\end{align}
\end{subequations}
where $\bar{T}_i^n(\bs{X})$, $\bar{M}_i^n(\bs{X})$, and $\bar{G}_i^n(\bs{X})$ are the components of the boundary tractions at $t^n$, $\bar{T}_i(\bs{X},t^n)$, $\bar{M}_i(\bs{X},t^n)$, and $\bar{G}_i(\bs{X},t^n)$, respectively.  

\section{Analysis}\label{S:analysis}
In this section we prove mass conservation, unconditional stability, and second-order accuracy of the time-integration algorithm proposed in Sec.\ref{SS:numerical_formulation_time}.  

    \subsection{Mass conservation}\label{SS:analysis_massconservation}
    Provided that Eqns.\eqref{E:cmuu_space_time} are satisfied for all $q^h(\bs{X})\in\mathcal{T}^h$, $\nu^h(\bs{X})\in\mathcal{T}^h$, and $\bs{w}^h(\bs{X})\in\mathcal{V}_w^h$, those equations are necessarily satisfied when we substitute the following for these test functions:
\begin{align}
q^h=1,\quad\nu^h=0,\quad\bs{w}^h=\bs{0}.\notag
\end{align}
One then readily obtains:
\begin{align}
\int_{\Omega}c^{h,n+1}\dif V=\int_{\Omega}c^{h,n}\dif V,\notag
\end{align}
which implies that mass is conserved from time $t^n$ to time $t^{n+1}$.

    \subsection{Stability}\label{SS:analysis_stability}
    We now investigate stability of the proposed time-integration algorithm.  
To this end, we assume that all Dirichlet and Neumann boundary conditions are time-independent; that is $\bar{u}_i$, $\bar{m}_i$, $\bar{g}_i$, $\bar{T}_i$, $\bar{M}_i$, and $\bar{G}_i$ are constant in time.  

Provided that Eqns. \eqref{E:cmuu_space_time} are satisfied for all $q^h(\bs{X})\in\mathcal{T}^h$, $\nu^h(\bs{X})\in\mathcal{T}^h$, and $\bs{w}^h(\bs{X})\in\mathcal{V}_w^h$, those equations are necessarily satisfied when we substitute the following for these test functions:
\begin{align}
q^h=\{\mu^h\}^n,\quad\nu^h=\left\{\!\frac{Dc^h}{Dt}\!\right\}^n,\quad\bs{w}^h=\frac{\bs{u}^{h,n+1}-\bs{u}^{h,n}}{\Delta t}.\notag
\end{align}
We then add the resulting three equations together and use identities \eqref{E:{mubar}deltac} and \eqref{E:{HWPB}delta} to obtain:
\begin{align}
\frac{\Pi^{h,n+1}-\Pi^{h,n}}{\Delta t}=-\int_{\Omega}\left(\{\mu^h\}^n_{,A}\{L_{AB}(c^h)\}^n\{\mu^h\}^n_{,B}+\frac{1}{24}\frac{\dif^3\bar{\mu}}{\dif c^3}(\xi)\frac{(\Delta c^h)^4}{\Delta t}
\right)\dif V,
\label{E:Pidiff}
\end{align}
where $\Pi^{h,n}$ is the space-time discrete total free energy at $t=t^n$ defined as:
\begin{align}
\Pi^{h,n}
&=\int_{\Omega}
\Psi_{c}(c^{h,n})
+\Psi_{s+e}(\bs{\zeta}^{h,n})\dif V
-\int_{\Gamma^T}u^{h,n}_i\bar{T}_i\dif S
-\int_{\Gamma^M}Du^{h,n}_i\bar{M}_i\dif S
-\int_{\Upsilon^G}u^{h,n}_i\bar{G}_i\dif C,\label{E:Pi^hn}
\end{align}
where spatial derivatives should be understood in the weak sense.  
Note that, as the mobility tensor is positive definite and $\dif^3\bar{\mu}/\!\!\dif c^3(\xi)$ is positive by definition \eqref{E:Psi_c}, the right-hand side of \eqref{E:Pidiff} is non-positive.  
Eqn. \eqref{E:Pidiff} therefore states that the discrete total free energy is non-increasing from time $t^n$ to time $t^{n+1}$ and the algorithm proposed in Sec.\ref{SS:numerical_formulation_time} is necessarily unconditionally stable.  
It should also be noted that, as seen in Eqn. \eqref{E:Pidiff}, numerical dissipation originates only from approximation of the logarithmic chemical potential $\{\bar{\mu}(c^h)\}^n$ and not from approximations $\{H(\bs{\zeta}^h)\}^n$, $\{W_A(\bs{\zeta}^h)\}^n$, $\{P_{iJ}(\bs{\zeta}^h)\}^n$, and $\{B_{iJK}(\bs{\zeta}^h)\}^n$; our multivariate Taylor expansion method applied to a multivariate polynomial free energy produces no numerical dissipation.  

    \subsection{Consistency and second-order accuracy}\label{SS:analysis_consistency}
    We proceed to show second-order accuracy of the proposed scheme.  
Following the standard treatment for the consistency analysis, we replace $c^{h,n}(\bs{X})$, $\mu^{h,n}(\bs{X})$, $\bs{u}^{h,n}(\bs{X})$, $c^{h,n+1}(\bs{X})$, $\mu^{h,n+1}(\bs{X})$, and $\bs{u}^{h,n+1}(\bs{X})$ in the time-discrete formulations \eqref{E:c_space_time}, \eqref{E:mu_space_time}, and \eqref{E:u_space_time} with the corresponding solutions  to the time-continuous problem \eqref{E:cmuu_space} at $t^n$ and $t^{n+1}$;
we denote the left-hand sides of the resulting equations by $I_c^n$, $I_\mu^n$, and $I_u^n$, respectively.  
The following approximations for an arbitrary function $\phi(\bs{X},t)$ readily obtained by Taylor expansions are of use:
\begin{align}
&\frac{\phi(\bs{X},t^{n+1})+\phi(\bs{X},t^n)}{2}=\phi(\bs{X},t^n)+\frac{\Delta t}{2}\frac{D\phi}{Dt}(\bs{X},t^n)+O(\Delta t^2),\notag\\
&\frac{\phi(\bs{X},t^{n+1})-\phi(\bs{X},t^n)}{\Delta t}=\frac{D\phi}{Dt}(\bs{X},t^n)+\frac{\Delta t}{2}\frac{D^2\phi}{Dt^2}(\bs{X},t^n)+O(\Delta t^2),\notag\\
&\Delta\phi(\bs{X}):=\phi(\bs{X},t^{n+1})-\phi(\bs{X},t^n)=\frac{D\phi}{Dt}(\bs{X},t^n)\Delta t+O(\Delta t^2).\notag
\end{align}
The definitions of the high-order terms $R^{c}$, $R^{\nabla c}_A$, $R^{F}_{iJ}$, and $R^{\nabla F}_{iJK}$ in \eqref{E:Rs} show them to be $O(\Delta t^2)$. Therefore,
one can, for instance, show that $\{P_{iJ}(\bs{\zeta}^h)\}^n$ given in \eqref{E:{P}} can after substitution of the time-continuous solutions be approximated as:
\begin{align}
\{P_{iJ}(\bs{\zeta}^h)\}^n=\tilde{P}_{iJ}(\bs{X},t^n)+\frac{\Delta t}{2}\frac{D\tilde{P}_{iJ}}{Dt}(\bs{X},t^n)+O(\Delta t^2),\notag
\end{align}
where $\tilde{P}_{iJ}(\bs{X},t):=P_{iJ}\left(\bs{\zeta}^h(\bs{X},t)\right)$.  
Treating other terms similarly, we can readily show the following:
\begin{subequations}
\begin{align}
I_c^n&=I_c(t^n)+\frac{\Delta t}{2}\frac{\dif I_c}{\dif t}(t^n)+O(\Delta t^2),\\
I_\mu^n&=I_\mu(t^n)+\frac{\Delta t}{2}\frac{\dif I_\mu}{\dif t}(t^n)+O(\Delta t^2),\\
I_u^n&=I_u(t^n)+\frac{\Delta t}{2}\frac{\dif I_u}{\dif t}(t^n)+O(\Delta t^2),
\end{align}
\label{E:I_cmuu^n}%
\end{subequations}
where $I_c\left(t\right)$, $I_\mu\left(t\right)$, and $I_u\left(t\right)$ are, respectively, the left-hand sides of Eqns. \eqref{E:c_space}, \eqref{E:mu_space}, and \eqref{E:u_space}.  
Since the first two terms of the right-hand side of each equation in \eqref{E:I_cmuu^n} are zero, one concludes that the proposed time-integration scheme \eqref{E:cmuu_space_time} is of order 2.  

We note here that in the above consistency analysis the specific formulas for $R^{c}$, $R^{\nabla c}_A$, $R^{F}_{iJ}$, and $R^{\nabla F}_{iJK}$ are unimportant.  
Indeed one can ignore some or all high-order terms existing in $R^{c}$, $R^{\nabla c}_A$, $R^{F}_{iJ}$, and $R^{\nabla F}_{iJK}$ when evaluating \eqref{E:{HWPB}}, with the resulting \emph{reduced formulations} remaining second-order accurate.  
Such reduced formulations lose unconditional stability, but are often equipped with practical accuracy.  
Requiring less computation, they can serve as good alternatives to the original formulation in many problems; see Sec.\ref{SS:numerical_example_reducedformulation} for examples.  

\section{Numerical examples}\label{S:numerical_example}
    In this section we demonstrate the robustness and accuracy of the numerical formulation proposed in Sec. \ref{S:numerical_formulation}.  
Parameters for the chemical free energy density function $\Psi_c$ introduced in Eqn. \eqref{E:Psi_c} are set as $A_1=1$ and $A_2=3$ 
so that this function takes its maximum at $c=0.5000$ and minimum around $c=0.0707$ and $0.9293$.  

\begin{figure}[b]
\begin{center}
        \includegraphics[scale=1]{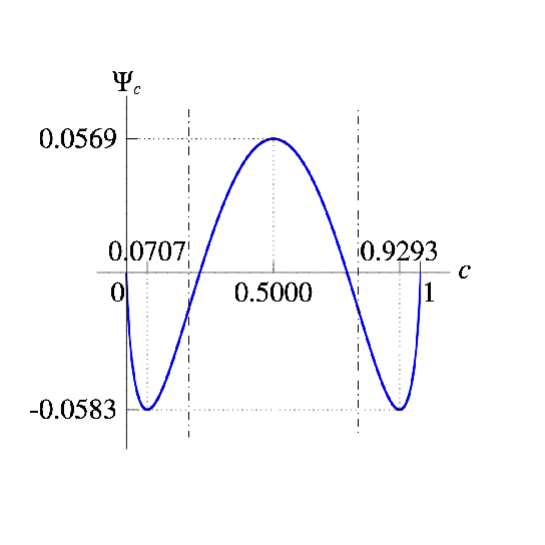}
        \caption{The chemical free energy density function $\Psi_c$, plotted against chemical composition $c$ for parameters used in our example problems.  }
        \label{Fi:Psi_c}
\end{center}
\end{figure}

\begin{figure}
\begin{center}
        \begin{subfigure}[b]{5.5cm}
                \centering
                \includegraphics[scale=1]{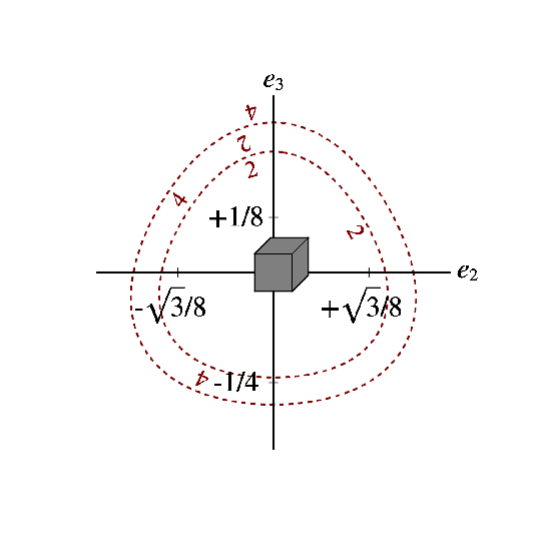}
                \caption{}
                \label{Fi:3well_00707}
        \end{subfigure}
        ~
        \begin{subfigure}[b]{5.5cm}
                \centering
                \includegraphics[scale=1]{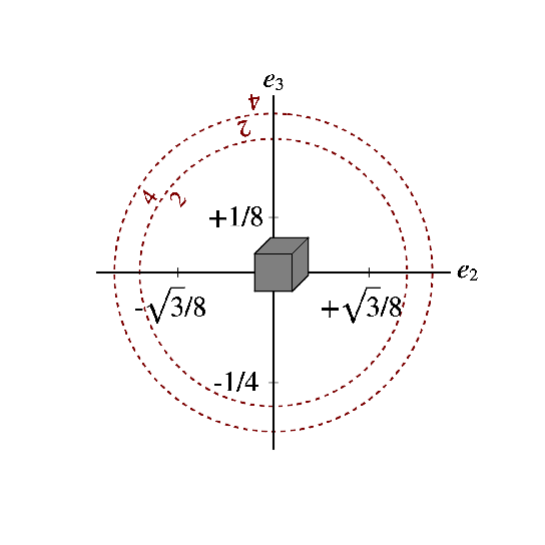}
                \caption{}
                \label{Fi:3well_05000}
        \end{subfigure}
        ~
        \begin{subfigure}[b]{5.5cm}
                \centering
                \includegraphics[scale=1]{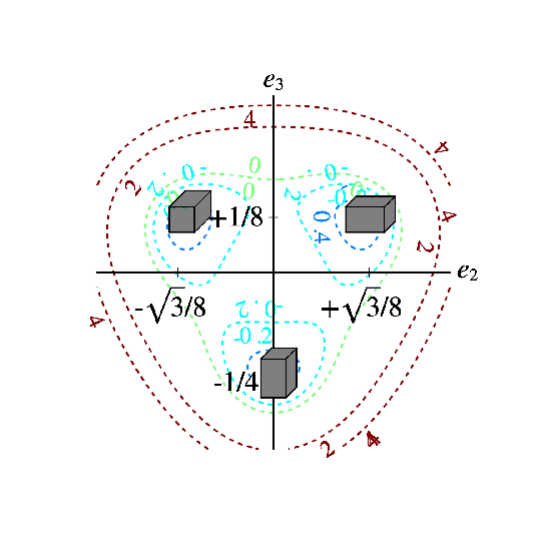}
                \caption{}
                \label{Fi:3well_09293}
        \end{subfigure}%
        \caption{Contour plots of the projection of the mechanical free energy density function $\Psi_e$ onto the $e_2-e_3$ plane for (\subref{Fi:3well_00707}) $c=0.0707$, (\subref{Fi:3well_05000}) $c=0.5000$, and (\subref{Fi:3well_09293}) $c=0.9293$.  The corresponding stable crystal structures are also depicted.  }
        \label{Fi:3well}
\end{center}
\end{figure}

Fig. \ref{Fi:Psi_c} shows plots of $\Psi_c$ against chemical composition $c$, where the chemical spinodal region is also shown as the interval between two dashed lines.  
We denote by $d(=0.1152)$ the difference between the maximum and minimum values of $\Psi_c$.  
The diffusion and mobility tensors appearing in Eqn. \eqref{E:Psi_s} and in Eqn. \eqref{E:c_nonequil_strong}, respectively, are defined as $K_{AB}(c)=2^{-6}\delta_{AB}$ and $L_{AB}(c)=6c(1-c)\delta_{AB}$.  
Parameters for the mechanical free energy density function $\Psi_e$ defined in Eqn.\eqref{E:Psi_e} are given as 
$e_{\textrm{chem}}=-1/16(c-0.48)$, $B_1(c)=13/4 r $, $B_2(c)=-5/32 r (c-0.50)$, $B_3(c)=1/4 r (c-0.50)$, $B_4(c)= r $, $B_5(c)=2 r $, and $B_6(c)=\!2^{-14} r  c$, where $ r =2^{11}d$. These free energy density function parameters have been chosen to produce a suitable microstructure; a wide range of microstructures results by variation of these parameters.
The free energy density component $\Psi_e$ thus defined characterizes crystallographic structural changes between a cubic phase and three energetically-equivalent tetragonal phases; the cubic phase loses stability and transforms into one of three stable tetragonal phases as the local composition $c$ increases.    
This transformation is represented by the projection of $\Psi_e$ onto the $e_2-e_3$ plane, showing a continuous transition from convex to non-convex form with \emph{three-wells} on the $e_2-e_3$ plane for $c>0.5$; see Fig. \ref{Fi:3well}.  
The distance on the $e_2-e_3$ plane from the origin to these three minima is designed to be $1/4$ when $c=1$; that is, minima occur at $(+\sqrt{3}/8,+1/8)$, $(-\sqrt{3}/8,+1/8)$, and $(0,-1/4)$, each corresponding to a tetragonal crystal structure that is elongated in the $X_1$-, $X_2$-, or $X_3$- direction, respectively.  
Stable crystal structures are also depicted in Fig. \ref{Fi:3well}.

We also note here that, to compute $\{H(\bs{\zeta}^h)\}^n$, $\{W_A(\bs{\zeta}^h)\}^n$, $\{P_{iJ}(\bs{\zeta}^h)\}^n$, and $\{B_{iJK}(\bs{\zeta}^h)\}^n$ defined in \eqref{E:{HWPB}} and \eqref{E:Rs}, one needs to take derivatives of $\Psi_{s+e}(\bs{\zeta})$ with respect to $c$, $c_A$, $F_{iJ}$, and $F_{iJ,K}$, respectively, $2$, $2$, $8$, and $2$ times. This follows from Equations (\ref{E:Psi_c})--(\ref{E:Psi_e}) and (\ref{E:ea})--(\ref{E:ef}). It is therefore sufficient to set $\kappa_c\leq 2$, $\kappa_{\nabla c}\leq 2$, $\kappa_{F}\leq 8$, and $\kappa_{\nabla F}\leq 2$.  

The reference domain occupies a unit cube $\Omega=(0,1)^3$ and
we set boundary conditions as displacement $u_i=0$ and displacement normal gradient $Du_i=0$ on $X_3=0$, traction $\bar{T}_i=0$ and higher-order traction $\bar{M}_i=0$ on $X_3=1$, and normal displacement $u_kN_k=0$, shear traction $\bar{T}_i-\bar{T}_kN_kN_i=0$, and higher-order traction $\bar{M}_i=0$ on $X_1,X_2=\{0,1\}$.  
Further, we set the higher-order traction jump $G_i=0$ on all edges where $u_i=0$ is not specified.  

In this work we employ isogeometric analysis (IGA) that allows for the easy construction of $C^p$-continuous basis functions for arbitrary order $p$.  
IGA has previously been adopted to treat higher-order spatial derivatives, e.g., in \cite{Gomez2008,Rudraraju2014} and we refer the reader to these works for details on the development of IGA basis functions with the desired degree of continuity.  
Our three-dimensional IGA basis functions are formed by the tensor products of one-dimensional, second-order B-spline basis functions on uniformly spaced \emph{knots}.  
The initial condition of the local composition $c$ is produced on a coarse $2^3$ mesh that has 2 \emph{elements}, or 4 basis functions, in each direction.  
The primitive one-dimensional B-spline basis functions in the $X_1$-, $X_2$-, and $X_3$-directions are indexed as $i_1$ ,$i_2$, and $i_3$, where $i_1, i_2, i_3=0,1,2,3$, and \emph{control points} for $i_1,i_2,i_3=1,2$ are given as $0.48+0.01\sin(999\sin(997i_1+991i_2+983i_3+1))$ and those for $i_1,i_2,i_3=0,3$ are computed to satisfy the boundary condition \eqref{E:c_equil_strong_bc}.  
This initial condition is projected onto finer meshes ($16^3$, $32^3$, and $64^3$ meshes) \emph{exactly} by successive uniform \emph{h-refinements} by \emph{knot-insertions} \cite{Cottrell2009}.  
Finally, initial conditions for the chemical potential and the displacement field are given as $\mu=0$ and $u_i=0$.  

We solve these three-dimensional, mechano-chemical, initial and boundary value problems using the unconditionally stable, second-order accurate time-integration algorithm proposed in Sec.\ref{S:numerical_formulation} with absolute residual error tolerance set to $10^{-10}$.  
Our open source research code \cite{mechanoChemistryStabilityAnalysisCode} is written in \textsf{C}, uses \textsf{PETSc 3.5} for linear/nonlinear solvers and \textsf{mathgl 2.3} for plotting.  We also utilized \textsf{Mathematica 10} to compute high-order indicial sums appearing in \eqref{E:{HWPB}} and \eqref{E:Rs}, and tangent matrices required for nonlinear solvers.

    \subsection{Temporal accuracy}\label{SS:numerical_example_time}
    In this section we study the temporal accuracy of our proposed formulation using a $32^3$ mesh throughout.  
We use progressively finer timesteps $\Delta t=4\!\times\! 10^{-3}$, $2\!\times\! 10^{-3}$, $1\!\times\! 10^{-3}$, $5\!\times\! 10^{-4}$, and $2.5\!\times\! 10^{-4}$ and integrate up to $t=4$, at which time the solutions were found to be almost at steady state.  

\begin{figure}
    \begin{center}
        \begin{tabular}{rp{15cm}}
             &
            \begin{tabular}{p{2.5cm}p{2.5cm}p{2.5cm}p{2.5cm}p{2.5cm}p{2.5cm}}
                \hspace{0.7cm}$t\!=\!0.08$ & 
                \hspace{0.7cm}$t\!=\!0.10$ &
                \hspace{0.7cm}$t\!=\!1.00$ &
                \hspace{0.7cm}$t\!=\!2.00$ &
                \hspace{0.7cm}$t\!=\!4.00$ &
                \vspace{0.5\baselineskip}
            \end{tabular}  \\
            \parbox[t]{2cm}{ $\Delta t\!=\!4\times\! 10^{-3}$} &
            \begin{tabular}{p{2.5cm}p{2.5cm}p{2.5cm}p{2.5cm}p{2.5cm}p{2.5cm}}
                \includegraphics[scale=1]{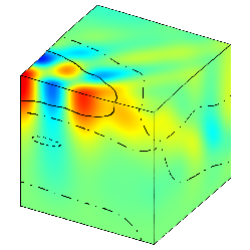} & 
                \includegraphics[scale=1]{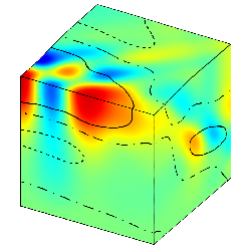} &
                \includegraphics[scale=1]{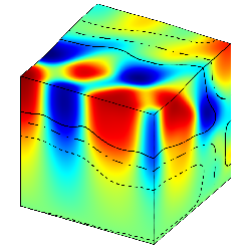} &
                \includegraphics[scale=1]{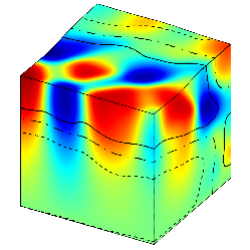} &
                \includegraphics[scale=1]{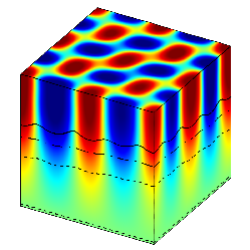} &
                
            \end{tabular}  \\
            \parbox[t]{2cm}{ $\Delta t\!=\!2\times\! 10^{-3}$} &
            \begin{tabular}{p{2.5cm}p{2.5cm}p{2.5cm}p{2.5cm}p{2.5cm}p{2.5cm}}
                \includegraphics[scale=1]{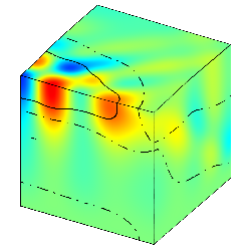} & 
                \includegraphics[scale=1]{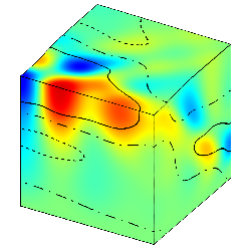} &
                \includegraphics[scale=1]{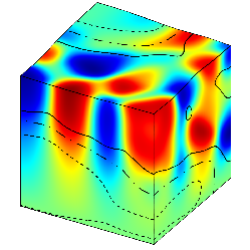} &
                \includegraphics[scale=1]{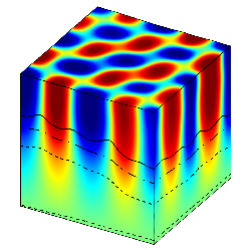} &
                \includegraphics[scale=1]{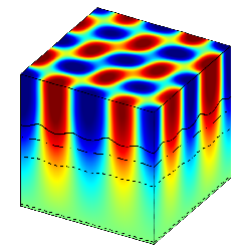} &
                \includegraphics[scale=1]{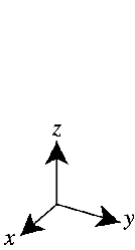}
            \end{tabular}  \\
            \parbox[t]{2cm}{ $\Delta t\!=\!1\times\! 10^{-3}$} &
            \begin{tabular}{p{2.5cm}p{2.5cm}p{2.5cm}p{2.5cm}p{2.5cm}p{2.5cm}}
                \includegraphics[scale=1]{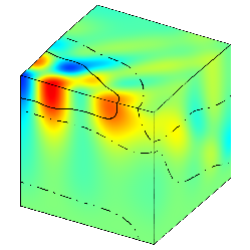} & 
                \includegraphics[scale=1]{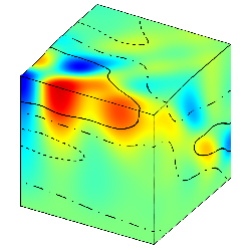} &
                \includegraphics[scale=1]{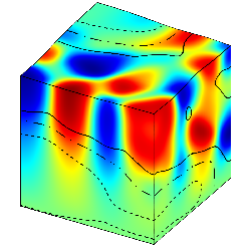} &
                \includegraphics[scale=1]{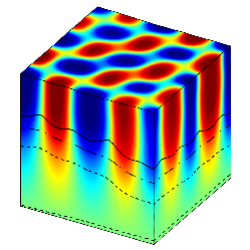} &
                \includegraphics[scale=1]{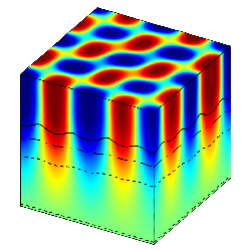} &
                \includegraphics[scale=1]{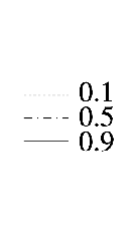}
            \end{tabular}  \\
            \parbox[t]{2cm}{ $\Delta t\!=\!5\times\! 10^{-4}$} &
            \begin{tabular}{p{2.5cm}p{2.5cm}p{2.5cm}p{2.5cm}p{2.5cm}p{2.5cm}}
                \includegraphics[scale=1]{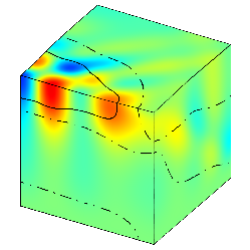} & 
                \includegraphics[scale=1]{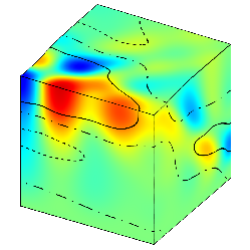} &
                \includegraphics[scale=1]{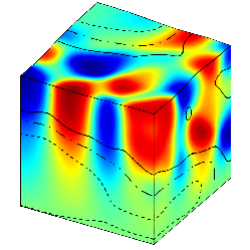} &
                \includegraphics[scale=1]{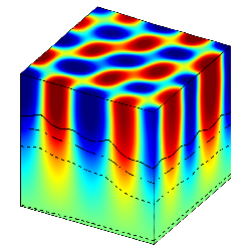} &
                \includegraphics[scale=1]{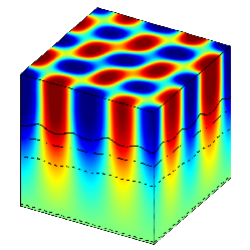} &
                \includegraphics[scale=1]{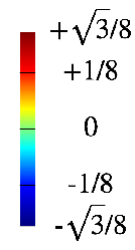} 
            \end{tabular}
        \end{tabular}
    \end{center}
    \caption{Plots of $e_2$ and contour curves of $c$ on deformed configurations at select times for solutions corresponding to four different time steps $\Delta t=4\!\times\! 10^{-3}$, $2\!\times\! 10^{-3}$, $1\!\times\! 10^{-3}$, and $5\!\times\! 10^{-4}$ on a $32^3$ mesh.  }
    \label{Fi:e1_c_time}
\end{figure}

Fig. \ref{Fi:e1_c_time} shows color plots of $e_2$ along with contour curves of $c$ for solutions obtained using $\Delta t=4\!\times\! 10^{-3}$, $2\!\times\! 10^{-3}$, $1\!\times\! 10^{-3}$, and $5\!\times\! 10^{-4}$.   
While $\Delta t=4\!\times\! 10^{-3}$ leads to a completely different morphological evolution implying insufficient temporal resolution, the
absence of visible differences between the solutions for $\Delta t=2\!\times\! 10^{-3}$, $1\!\times\! 10^{-3}$, and $5\!\times\! 10^{-4}$ indicates convergence of the microstructure as timesteps are refined.  

\begin{figure}
\begin{center}
        \begin{subfigure}[b]{5.5cm}
                \centering
                \includegraphics[scale=1]{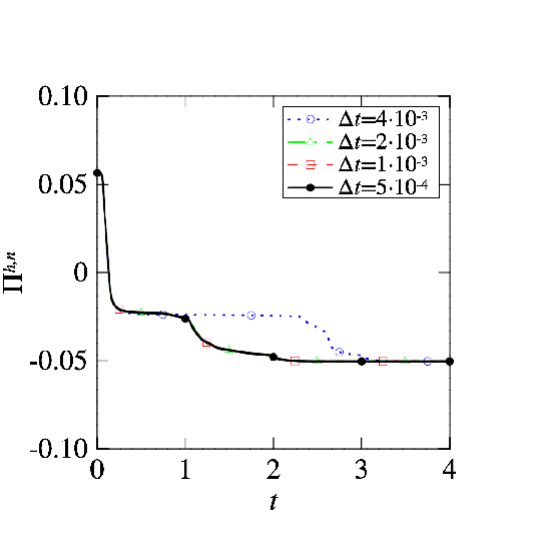}
                \caption{}
                \label{Fi:time_energy_time_normal}
        \end{subfigure}
        ~
        \begin{subfigure}[b]{5.5cm}
                \centering
                \includegraphics[scale=1]{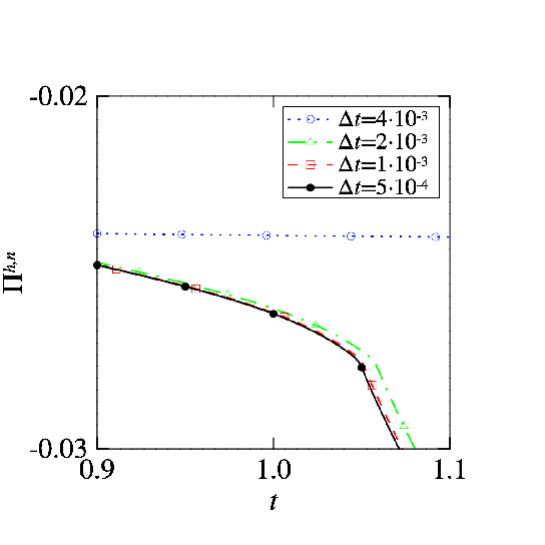}
                \caption{}
                \label{Fi:time_energy_time_zoom}
        \end{subfigure}
        ~
        \begin{subfigure}[b]{5.5cm}
                \centering
                \includegraphics[scale=1]{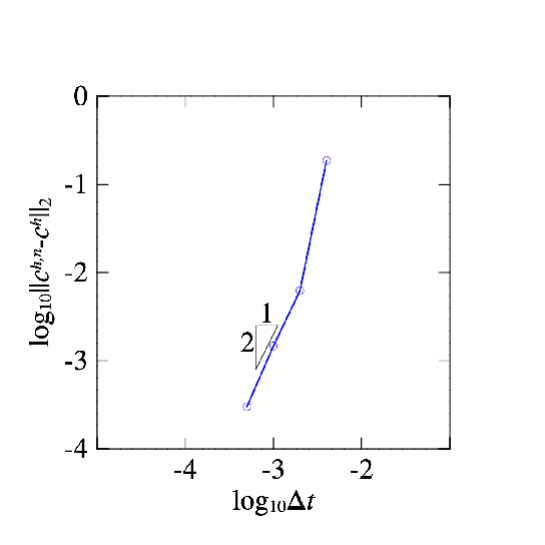}
                \caption{}
                \label{Fi:time_energy_time_order}
        \end{subfigure}%
        \caption{Plots of discrete total free energy $\Pi^{h,n}$ against time $t$ for (\subref{Fi:time_energy_time_normal}) $t\in[0,4]$ and (\subref{Fi:time_energy_time_zoom}) $t\in[0.9,1.1]$ for four different timesteps $\Delta t=4\!\times\! 10^{-3}$, $2\!\times\! 10^{-3}$, $1\!\times\! 10^{-3}$, and $5\!\times\! 10^{-4}$.  (\subref{Fi:time_energy_time_order}) Plots of error, $\norm{c^{h,n}-c^{h}}_2$, at $t=1$ against timestep $\Delta t$ on a logarithmic scale.  A fixed $32^3$ mesh was used.  }
        \label{Fi:time_energy_time}
\end{center}
\end{figure}

Figs. \ref{Fi:time_energy_time}(\subref{Fi:time_energy_time_normal}) and \ref{Fi:time_energy_time}(\subref{Fi:time_energy_time_zoom}) show plots of discrete total free energy $\Pi^{h,n}$ against time for $t\in[0,4]$ and $t\in[0.9,1.1]$, respectively, for solutions corresponding to $\Delta t=4\!\times\! 10^{-3}$, $2\!\times\! 10^{-3}$, $1\!\times\! 10^{-3}$, and $5\!\times\! 10^{-4}$.  
One can observe convergence of the free energies of the solutions on $t\in[0,4]$ with time step refinement; we also remark that, while the numerical solution with $\Delta t = 4\times 10^{-3}$ has a markedly different microstructure from those computed with finer timesteps (Fig. \ref{Fi:e1_c_time}), it does indeed converge to the same free energy at the steady state (Fig. \ref{Fi:time_energy_time}(\subref{Fi:time_energy_time_normal})).  
Of note is that the discrete free energy is non-increasing for any time step, implying unconditional stability of our proposed time-discrete formulation as expected from the analysis in Sec. \ref{SS:analysis_stability}.   
In this regard we point out that, though the scheme is unconditionally stable, the time-step size still sets a constraint on the solvability of the nonlinear system; the nonlinear solver quit converging to the set tolerance around $t=0.088$ when we used $\Delta t=8\!\times\! 10^{-3}$, which is in any case too large to have a reasonable solution for this specific problem.  

Finally, we regard the solution for $\Delta t=2.5\!\times\! 10^{-4}$ as exact and compute $L^2$-norm of the error of the solution field $c$, $\norm{c^{h,n}-c^{h}}_2$, at $t=1$ for each solution.  
Fig. \ref{Fi:time_energy_time}(\subref{Fi:time_energy_time_order}) shows plots of error $\norm{c^{h,n}-c^{h}}_2$ against time step $\Delta t$ on a logarithmic scale, where one observes a second-order temporal convergence as expected from the analysis in Sec. \ref{SS:analysis_consistency}.  

    \subsection{Spatial convergence}\label{SS:numerical_example_space}
\begin{figure}
    \begin{center}
        \begin{tabular}{rp{15cm}}
             &
            \begin{tabular}{p{2.5cm}p{2.5cm}p{2.5cm}p{2.5cm}p{2.5cm}p{2.5cm}}
                \hspace{0.7cm}$t\!=\!0.08$ & 
                \hspace{0.7cm}$t\!=\!0.10$ &
                \hspace{0.7cm}$t\!=\!1.00$ &
                \hspace{0.7cm}$t\!=\!2.00$ &
                \hspace{0.7cm}$t\!=\!4.00$ &
                \vspace{0.5\baselineskip}
            \end{tabular}  \\
            \parbox[t]{2cm}{ $16^3$ mesh} &
            \begin{tabular}{p{2.5cm}p{2.5cm}p{2.5cm}p{2.5cm}p{2.5cm}p{2.5cm}}
                \includegraphics[scale=1]{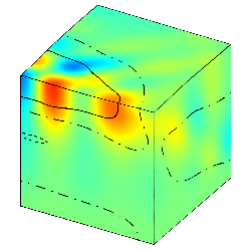} & 
                \includegraphics[scale=1]{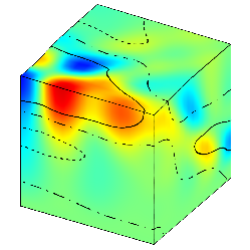} &
                \includegraphics[scale=1]{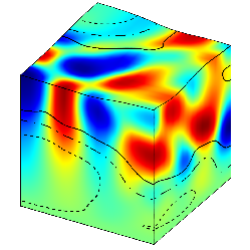} &
                \includegraphics[scale=1]{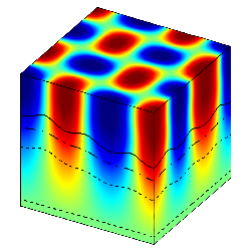} &
                \includegraphics[scale=1]{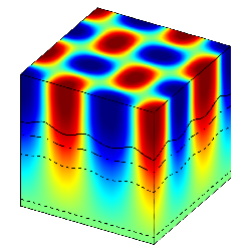} &
                
            \end{tabular}  \\
            \parbox[t]{2cm}{ $32^3$ mesh} &
            \begin{tabular}{p{2.5cm}p{2.5cm}p{2.5cm}p{2.5cm}p{2.5cm}p{2.5cm}}
                \includegraphics[scale=1]{./plot_608_e1_000080.eps} & 
                \includegraphics[scale=1]{./plot_608_e1_000100.eps} &
                \includegraphics[scale=1]{./plot_608_e1_001000.eps} &
                \includegraphics[scale=1]{./plot_608_e1_002000.eps} &
                \includegraphics[scale=1]{./plot_608_e1_004000.eps} &
                \includegraphics[scale=1]{./plot_xyz.eps}
            \end{tabular}  \\
            \parbox[t]{2cm}{ $64^3$ mesh} &
            \begin{tabular}{p{2.5cm}p{2.5cm}p{2.5cm}p{2.5cm}p{2.5cm}p{2.5cm}}
                \includegraphics[scale=1]{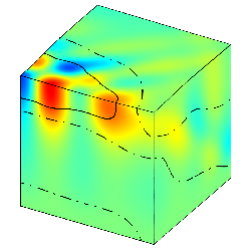} & 
                \includegraphics[scale=1]{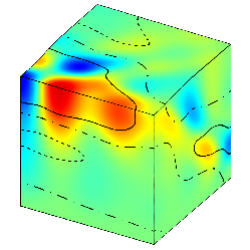} &
                \includegraphics[scale=1]{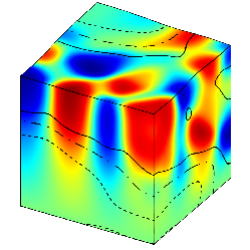} &
                \includegraphics[scale=1]{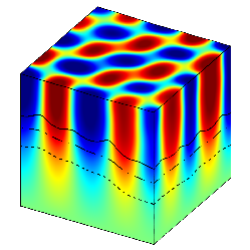} &
                \includegraphics[scale=1]{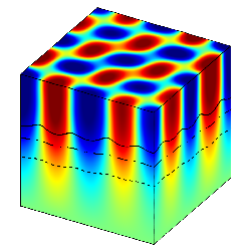} &
                \includegraphics[scale=1]{./plot_legend.eps}
            \end{tabular}  \\
            \parbox[t]{2cm}{ $64^3$ mesh\\ $\Delta t=1\!\times\!10^{-3}$} &
            \begin{tabular}{p{2.5cm}p{2.5cm}p{2.5cm}p{2.5cm}p{2.5cm}p{2.5cm}}
                \includegraphics[scale=1]{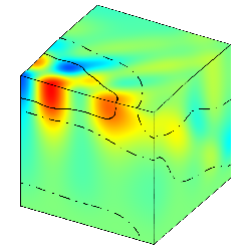} & 
                \includegraphics[scale=1]{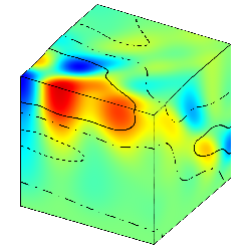} &
                \includegraphics[scale=1]{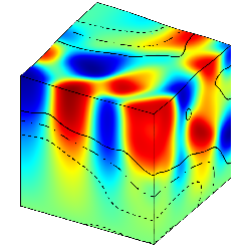} &
                \includegraphics[scale=1]{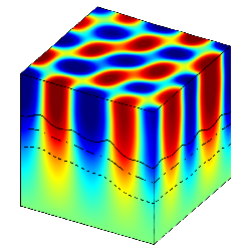} &
                \includegraphics[scale=1]{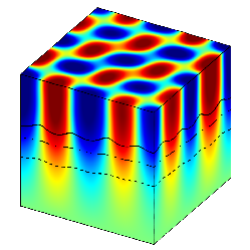} &
                \includegraphics[scale=1]{./plot_colorbar.eps} 
            \end{tabular}
        \end{tabular}
    \end{center}
    \caption{Plots of $e_2$ and contour curves of $c$ on deformed configurations at select times for solutions computed on three different meshes, $16^3$, $32^3$ and $64^3$ with $\Delta t=2\!\times\! 10^{-3}$.  The solution on a $64^3$ mesh with $\Delta t=1\!\times\! 10^{-3}$ is also shown for comparison.  }
    \label{Fi:e1_c_space}
\end{figure}

\begin{figure}
\begin{center}
        \begin{subfigure}[b]{5.5cm}
                \centering
                \includegraphics[scale=1]{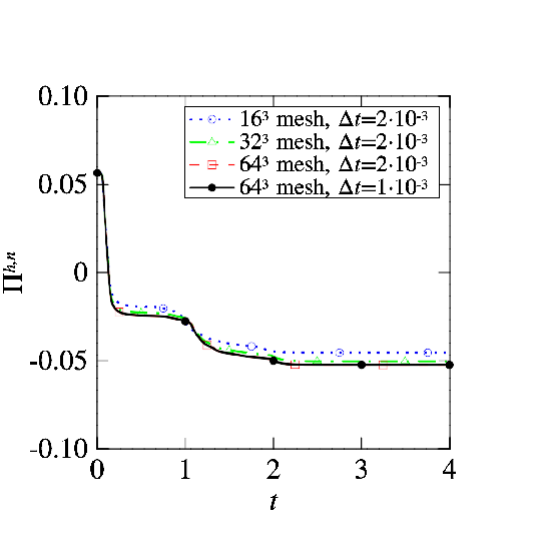}
                \caption{}
                \label{Fi:time_energy_space_normal}
        \end{subfigure}
        ~
        \begin{subfigure}[b]{5.5cm}
                \centering
                \includegraphics[scale=1]{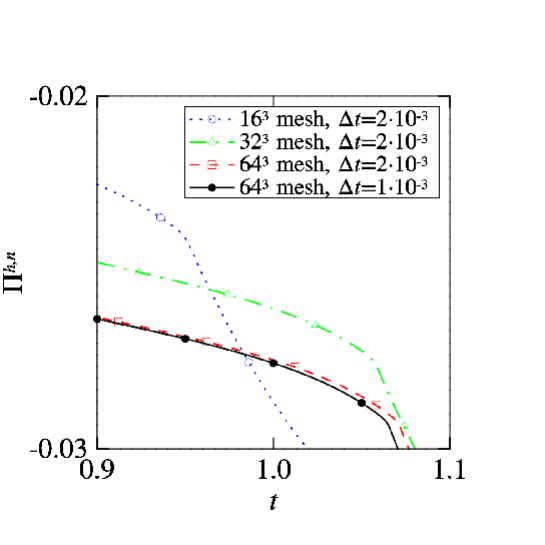}
                \caption{}
                \label{Fi:time_energy_space_zoom}
        \end{subfigure}%
        \caption{Plots of discrete total free energy $\Pi^{h,n}$ against time $t$ for (\subref{Fi:time_energy_space_normal}) $t\in[0,4]$ and (\subref{Fi:time_energy_space_zoom}) $t\in[0.9,1.1]$ for three different meshes, $16^3$, $32^3$, and $64^3$, with timestep $\Delta t=2\!\times\! 10^{-3}$.  The solution on a $64^3$ mesh with $\Delta t=1\!\times\! 10^{-3}$ is also shown for comparison.  }
        \label{Fi:time_energy_space}
\end{center}
\end{figure}

    We continue in this section to investigate spatial convergence of the solutions with mesh refinement.  
We solve problems on three different meshes, $16^3$, $32^3$, and $64^3$, using a fixed timestep $\Delta t=2\!\times\! 10^{-3}$ that is regarded as small enough.  
Time-integration is again performed up to $t=4$.  
Fig. \ref{Fi:e1_c_space} shows the temporal evolution of the microstructure on these three meshes.  
One observes that, while the $16^3$ mesh seems under-resolved, no further morphological changes appear under refinement from $32^3$ mesh to $64^3$ mesh, giving good evidence of spatial convergence of the microstructure.  
The solution on a $64^3$ mesh with $\Delta t=1\!\times\!10^{-3}$ is also plotted to show that $\Delta t=2\!\times\!10^{-3}$ gives sufficient temporal resolution for this spatial convergence analysis.  
Figs. \ref{Fi:time_energy_space}(\subref{Fi:time_energy_space_normal}) and \ref{Fi:time_energy_space}(\subref{Fi:time_energy_space_zoom}) show plots of corresponding discrete total free energy $\Pi^{h,n}$ against time for $t\in[0,4]$ and $t\in[0.9,1.1]$, respectively, where one can further observe convergence with respect to free energy of the numerical solutions under mesh refinement.  
Discrete total free energy plots for the solution on the $64^3$ mesh with $\Delta t=1\!\times\!10^{-3}$ are also shown, which provides further assurance that the timestep of $\Delta t=2\!\times\! 10^{-3}$ is small enough for this observation.  

\begin{figure}
\begin{center}
        \begin{subfigure}[b]{5.5cm}
                \centering
                \includegraphics[scale=1]{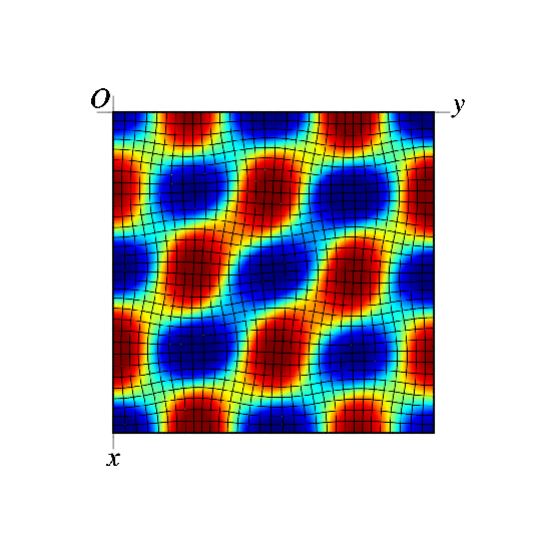}
                \caption{}
                \label{Fi:e2}
        \end{subfigure}
        ~
        \begin{subfigure}[b]{5.5cm}
                \centering
                \includegraphics[scale=1]{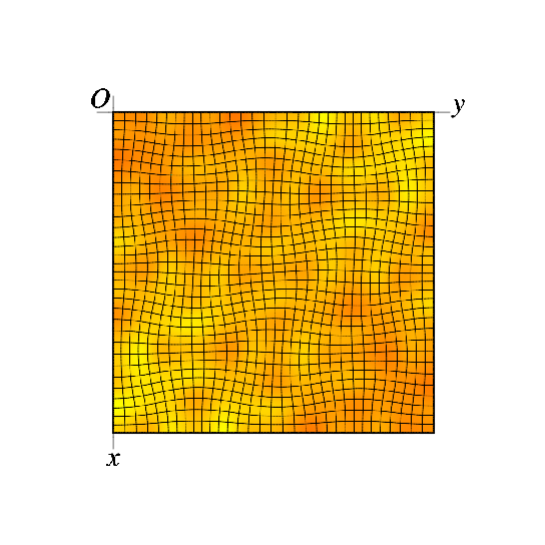}
                \caption{}
                \label{Fi:e3}
        \end{subfigure}
        ~
        \begin{subfigure}[b]{5.5cm}
                \centering
                \includegraphics[scale=1]{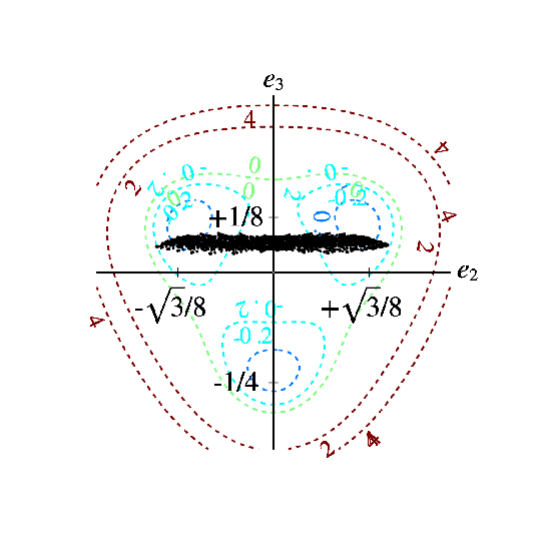}
                \caption{}
                \label{Fi:e2e3}
        \end{subfigure}%
        \caption{Plots of (\subref{Fi:e2}) $e_2$ and (\subref{Fi:e3}) $e_3$ on $Z=1$ obtained on the $64^3$ mesh with $\Delta t=2\!\times\! 10^{-3}$, overlaid with $32^2$ plotting meshes.  (\subref{Fi:e2e3}) Dot plots of $e_2$ and $e_3$ values on $X_3=1$ evaluated over a uniformly spaced square array of $64\times 64$ points, laid over the contour plots originally shown in Fig. \ref{Fi:3well}(\subref{Fi:3well_09293}).  }
        \label{Fi:e2_e3}
\end{center}
\end{figure}

We conclude this section by investigating the converged microstructure obtained in our numerical analysis.  
Figs. \ref{Fi:e2_e3}(\subref{Fi:e2}) and \ref{Fi:e2_e3}(\subref{Fi:e3}) show the top views of color plots of $e_2$ and $e_3$ in the deformed configuration at $t=4$ computed on a $64^3$ mesh with $\Delta t=2\times 10^{-3}$.  
For the sake of visualization of the deformation, these top views are overlaid with distorted $32^2$ meshes. The deformation reveals twin boundaries between two of the three tetragonal variants, viz. those two living in the upper-half plane in Fig. \ref{Fi:3well}(\subref{Fi:3well_09293}).  
Figs. \ref{Fi:e2_e3}(\subref{Fi:e2}) and \ref{Fi:e2_e3}(\subref{Fi:e3}) also make clear the large deformations that the microstructures suffer; note the distorted mesh corroborating the strain values in the legends of Figs. \ref{Fi:e1_c_time} and \ref{Fi:e1_c_space}.  
Fig. \ref{Fi:e2_e3}(\subref{Fi:e2e3}) shows dot plots of $(e_2,e_3)$ for a square array of $64\times 64$ points, uniformly spaced along $X_1$ and $X_2$ , and lying on the $X_3=1$ plane. Also reproduced, are the contour plots originally shown in Fig. \ref{Fi:3well}(\subref{Fi:3well_09293}).  
The sharp localization of the strain state in $e_2-e_3$  space is understood to be a consequence of the specific boundary conditions employed for this problem. We draw attention to the localization in the vicinity of only two of the three wells in $e_2-e_3$  space. The points lying between the two wells are found to lie in the twin boundaries between the corresponding variants on the physical domain $\Omega$. In other computations we have found the equidistribution of the strain state across all three wells when perfectly symmetric boundary conditions are used.

    \subsection{Reduced formulations}\label{SS:numerical_example_reducedformulation}
    In the preceding numerical examples, we have evaluated $\{\bar{\mu}(c^h)\}^n$, $\{H(\bs{\zeta}^h)\}^n$, $\{W_A(\bs{\zeta}^h)\}^n$, $\{P_{iJ}(\bs{\zeta}^h)\}^n$, and $\{B_{iJK}(\bs{\zeta}^h)\}^n$  following \eqref{E:{HWPB}} while using the full summations in
\eqref{E:Rs} given by $\kappa_c\leq 2$, $\kappa_{\nabla c}\leq 2$, $\kappa_{F}\leq 8$, and $\kappa_{\nabla F}\leq 2$.  
In this section we introduce reduced formulations in which some high-order terms in $\{\bar{\mu}(c^h)\}^n$, $\{H(\bs{\zeta}^h)\}^n$, $\{W_A(\bs{\zeta}^h)\}^n$, $\{P_{iJ}(\bs{\zeta}^h)\}^n$, and $\{B_{iJK}(\bs{\zeta}^h)\}^n$ are ignored; specifically, we consider two formulations obtained by setting $\kappa_{F}\leq 2$ and $\kappa_{F}\leq 4$ instead of $\kappa_{F}\leq 8$.  
These reduced formulations are \emph{not} unconditionally stable, but often provide solutions of sufficient quality at lower computational cost.  

To demonstrate this point we solve the example problem encountered in Sec. \ref{SS:numerical_example_time} on a $32^3$ mesh with $\Delta t=5\!\times\! 10^{-4}$ up to $t=4$ using these reduced formulations and compare them with the full formulation.  Using 2.60GHz Intel Xeon E5-2670 processors on $8\times 8\times 8$ partitions, the actual time required for the time-integration was measured for each formulation.  
The results are summarized in Table \ref{Ta:comput_time}.  
The reduced formulation with $\kappa_{F}\leq 2$ was about $2\times$ as fast as the full formulation with $\kappa_{F}\leq 8$, but the solution diverged around $t=1$, while the reduced formulation with $\kappa_{F}\leq 4$ was about $1.5\times$ as fast, retaining practical stability.  

\begin{figure}
    \begin{center}
        \begin{tabular}{rp{15cm}}
             &
            \begin{tabular}{p{2.5cm}p{2.5cm}p{2.5cm}p{2.5cm}p{2.5cm}p{2.5cm}}
                \hspace{0.7cm}$t\!=\!0.08$ & 
                \hspace{0.7cm}$t\!=\!0.10$ &
                \hspace{0.7cm}$t\!=\!1.00$ &
                \hspace{0.7cm}$t\!=\!2.00$ &
                \hspace{0.7cm}$t\!=\!4.00$ &
                \vspace{0.5\baselineskip}
            \end{tabular}  \\
            \parbox[t]{2cm}{$\kappa_{F}\leq 2$} &
            \begin{tabular}{p{2.5cm}p{2.5cm}p{2.5cm}p{2.5cm}p{2.5cm}p{2.5cm}}
                \includegraphics[scale=1]{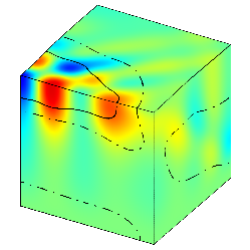} & 
                \includegraphics[scale=1]{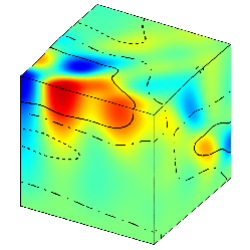} &
                \includegraphics[scale=1]{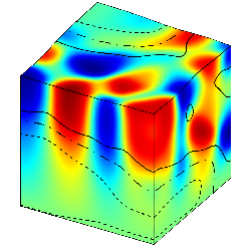} &
                &
                &
                \includegraphics[scale=1]{./plot_xyz.eps}
            \end{tabular}  \\
            \parbox[t]{2cm}{$\kappa_{F}\leq 4$} &
            \begin{tabular}{p{2.5cm}p{2.5cm}p{2.5cm}p{2.5cm}p{2.5cm}p{2.5cm}}
                \includegraphics[scale=1]{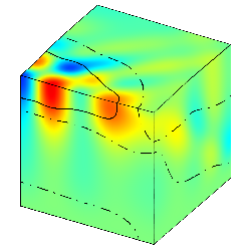} & 
                \includegraphics[scale=1]{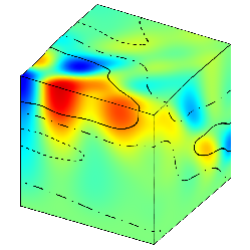} &
                \includegraphics[scale=1]{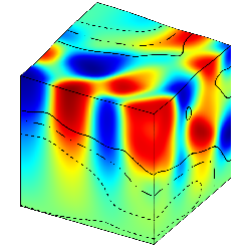} &
                \includegraphics[scale=1]{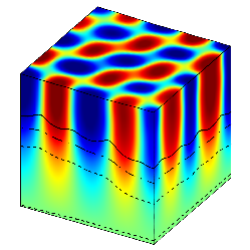} &
                \includegraphics[scale=1]{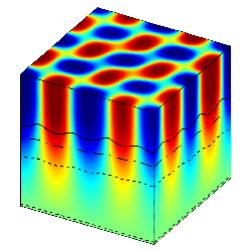} &
                \includegraphics[scale=1]{./plot_legend.eps}
            \end{tabular}  \\
            \parbox[t]{2cm}{$\kappa_{F}\leq 8$} &
            \begin{tabular}{p{2.5cm}p{2.5cm}p{2.5cm}p{2.5cm}p{2.5cm}p{2.5cm}}
                \includegraphics[scale=1]{./plot_610_e1_000080.eps} & 
                \includegraphics[scale=1]{./plot_610_e1_000100.eps} &
                \includegraphics[scale=1]{./plot_610_e1_001000.eps} &
                \includegraphics[scale=1]{./plot_610_e1_002000.eps} &
                \includegraphics[scale=1]{./plot_610_e1_004000.eps} &
                \includegraphics[scale=1]{./plot_colorbar.eps} 
            \end{tabular}
        \end{tabular}
    \end{center}
    \caption{Plots of $e_2$ and contour curves of $c$ on deformed configurations at select times for two reduced formulations, one with $\kappa_{F}\leq 2$ and the other one with $\kappa_{F}\leq 4$, and the full formulation with $\kappa_{F}\leq 8$.  Solutions were computed on a $32^3$ mesh with $\Delta t=5\!\times\! 10^{-4}$.  }
    \label{Fi:e1_c_reducedformulation}
\end{figure}

Fig. \ref{Fi:e1_c_reducedformulation} shows color plots of $e_2$ and contour curves of $c$ for these solutions at select times $t$, where
one can observe that microstructures obtained by the reduced formulations are almost identical to those obtained by the full formulation, although the reduced formulation with $\kappa_{F}\leq 2$ diverged around $t=1$.  

\begin{figure}
\begin{center}
        \begin{subfigure}[b]{5.5cm}
                \centering
                \includegraphics[scale=1]{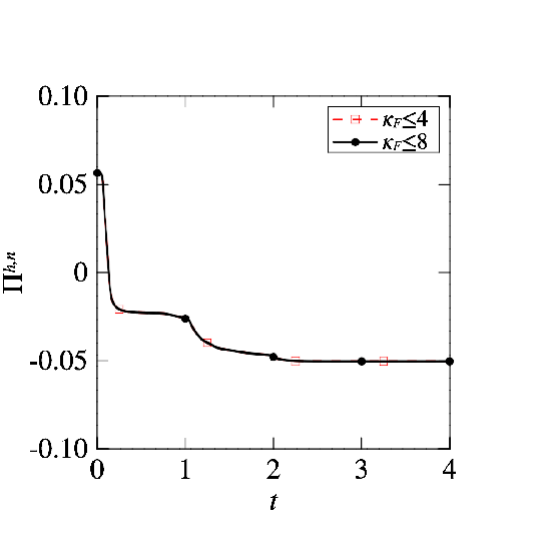}
                \caption{}
                \label{Fi:time_energy_reducedformulation_normal}
        \end{subfigure}
        ~
        \begin{subfigure}[b]{5.5cm}
                \centering
                \includegraphics[scale=1]{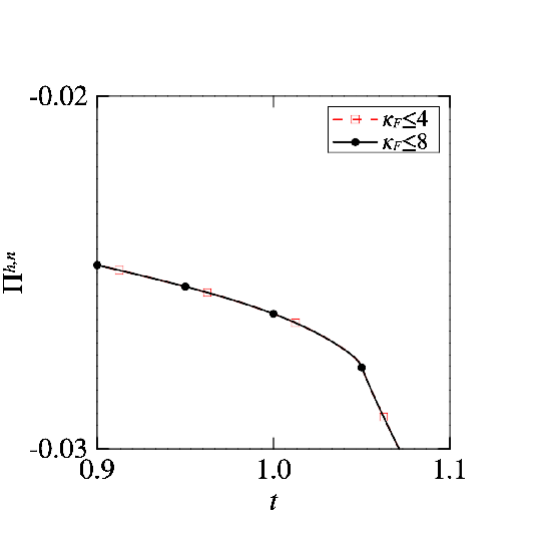}
                \caption{}
                \label{Fi:time_energy_reducedformulation_zoom}
        \end{subfigure}%
        \caption{Plots of discrete total free energy $\Pi^{h,n}$ against time $t$ for (\subref{Fi:time_energy_space_normal}) $t\in[0,4]$ and (\subref{Fi:time_energy_space_zoom}) $t\in[0.9,1.1]$ for two reduced formulations, one with $\kappa_{F}\leq 2$ and the other one with $\kappa_{F}\leq 4$, and the full formulation with $\kappa_{F}\leq 8$.  Solutions were computed on a $32^3$ mesh with $\Delta t=5\!\times\! 10^{-4}$.  }
        \label{Fi:time_energy_reducedformulation}
\end{center}
\end{figure}

Figs. \ref{Fi:time_energy_reducedformulation}(\subref{Fi:time_energy_reducedformulation_normal}) and \ref{Fi:time_energy_reducedformulation}(\subref{Fi:time_energy_reducedformulation_zoom}) show plots of corresponding discrete total free energy $\Pi^{h,n}$ against time for $t\in[0,4]$ and $t\in[0.9,1.1]$, respectively.  
The energy corresponding to the reduced formulation with $\kappa_{F}\leq 2$ takes slightly larger values than the other two formulations until it diverges in the neighborhood of $t=1$.
In contrast, the energy corresponding to the reduced formulation with $\kappa_{F}\leq 4$ is indistinguishable from that of the full formulation.  

This study indicates that, despite the possible loss of unconditional stability, reduced formulations can produce solutions that are of sufficient quality in practice and can serve as alternatives to the full formulation when faster time-integration is demanded.

\begin{remark}
Though the reduced formulation with $\kappa_{F}\leq 4$ may also seem to be unconditionally stable in the above problem, the authors believe otherwise; we indeed were able to observe apparent free energy oscillation of $O(10^{-13})$ after steady state is roughly achieved around $t=4$.  
This might be due to the numerical truncation error as the absolute residual error tolerance was set as $10^{-10}$ throughout the analysis, but we further note that the full formulation with $\kappa_{F}\leq 8$ always produced solutions of decreasing free energy up to the precision of $O(10^{-16})$ even with the same level of error tolerance.  
The other observation of possible interest is that, the time-step size tends to put severer constraints on the solvability of the nonlinear system for reduced formulations as more high-order terms are ignored; it was more apparent for the reduced formulation with $\kappa_{F}\leq 2$ than for the reduced formulation with $\kappa_{F}\leq 4$.  
In the authors' experience, however, those reduced formulations with $\kappa_{F}\leq 4$ and $\kappa_{F}\leq 2$ still show better performance in this regard than the conventional Backward Euler method.  
\end{remark}

\begin{table}
\begin{center}
\begin{tabular}[t]{|c|c|c|}
\hline
formulation & $t=1$ & $t=4$\\
\hline
$\kappa_{F}\leq 2$ & 12 & -\\
$\kappa_{F}\leq 4$ & 16 & 64\\
$\kappa_{F}\leq 8$ & 23 & 93\\
\hline
\end{tabular}
\end{center}
\caption{Time, in hour, required to compute solutions up to $t=1$ and $t=4$ on $32^3$ mesh with $\Delta t=5\!\times\!10^{-4}$ using 2.60GHz Intel Xeon E5-2670 processors on $8\times 8\times 8$ partitions for two reduced formulations, corresponding to $\kappa_{F}\leq 2$ and $\kappa_{F}\leq 4$, and the full formulation, corresponding to $\kappa_{F}\leq 8$.  }
\label{Ta:comput_time}
\end{table}

\section{Conclusion}\label{S:conclusion}
    We have developed a class of unconditionally stable, second-order accurate time-integration algorithms for nonlinear, mechano-chemical problems characterized by free energy functions that are non-convex in strain-composition space, and that must be stabilized by introducing spatial gradients of these fields. The associated phenomenology, which we term as mechano-chemical spinodal decomposition, includes the formation of microstructural features and transient phenomena. Their resolution by first-order schemes such as the Backward Euler algorithm does not necessarily preserve the free energy decay that is a consequence of the second law of thermodynamics. 

The approach presented here has wide applicability in the design of stable, second-order schemes for coupled problems of mechanics and transport. Its use hinging on the Taylor expansion, can, in principle, be extended to any free energy function that is a multivariate polynomial of direct components, component gradients, and higher-order component gradients. Such functions are guaranteed to have finite Taylor-series, and therefore, analytic forms, which are required of the constitutive relations to preserve unconditional stability.
While evaluation of the Taylor series expansions in the code comes at a cost, we have found that by exploiting symmetries inherent in the higher-order derivatives this cost can be substantially reduced. Furthermore, we have demonstrated that reduced formulations that truncate the higher-order terms in the Taylor series also can perform well for the initial and boundary value problems we have considered, even though unconditional stability can no longer be proven. \emph{Our analysis demonstrates, however, that second-order accuracy is maintained, regardless.} 

To the best of our knowledge, this is the first treatment presenting stable and second-order schemes for systems that incorporate Toupin's theory of gradient elasticity at finite strains. It has potential for extension to problems incorporating advection and reaction terms in the transport equation, as well as to systems that couple with the Allen-Cahn \cite{AllenCahn1979} treatment for evolution of non-conserved order parameters. It could thus cover a wide range of phase transformation phenomena involving solids as well as fluid phases in materials systems arising in battery, semi-conductor, polymer and structural applications.
 

\section*{Acknowledgments}
The mathematical and  numerical formulation, and the computations presented here have been carried out as part of research supported by the U.S. Department of Energy, Office of Basic Energy Sciences, Division of Materials Sciences and Engineering under Award \#DE-SC0008637 that funds the PRedictive Integrated Structural Materials Science (PRISMS) Center at University of Michigan.
\bibliographystyle{amsplain}
\bibliography{reference}

\providecommand{\bysame}{\leavevmode\hbox to3em{\hrulefill}\thinspace}
\providecommand{\MR}{\relax\ifhmode\unskip\space\fi MR }
\providecommand{\MRhref}[2]{%
  \href{http://www.ams.org/mathscinet-getitem?mr=#1}{#2}
}
\providecommand{\href}[2]{#2}
\begin{thebibliography}{10}

\bibitem{AllenCahn1979}
S.M. Allen and J.W Cahn, \emph{A microscopic theory for antiphase boundary
  motion and its application to antiphase boundary coarsening}, Acta
  Metallurgica \textbf{27} (1979), 1085--1091.

\bibitem{Boyer2011}
F.~Boyer and S~Minjeaud, \emph{{N}umerical schemes for a three component
  {C}ahn-{H}illiard model}, ESAIM.Mathematical Modelling and Numerical Analysis
  \textbf{45} (2011), 697--738.

\bibitem{CahnHilliard1958}
J.W Cahn and J.E. Hilliard, \emph{{F}ree energy of a nonuniform system. {I}.
  {I}nterfacial energy}, Journal of Chemical Physics \textbf{28} (1958),
  258--267.

\bibitem{Cottrell2009}
J.~Austin Cottrell, Thomas J.~R. Hughes, and Yuri Bazilevs,
  \emph{{I}sogeometric {A}nalysis}, John Wiley \& Sons, Ltd, 2009.

\bibitem{Du1991}
Qiang Du and R.~A. Nicolaides, \emph{{N}umerical analysis of a continuum model
  of phase transition}, SIAM Journal on Numerical Analysis \textbf{28} (1991),
  1310--1322.

\bibitem{Elliott1989a}
C.~M. Elliott, \emph{{M}athematical models for phase change problems},
  ch.~{T}he {C}ahn-{H}illiard Model for the Kinetics of Phase Separation,
  pp.~35--73, Birkh{\"a}user Basel, 1989.

\bibitem{Elliott1993}
C.~M. Elliott and A.~M. Stuart, \emph{{T}he global dynamics of discrete
  semilinear parabolic equations}, SIAM Journal on Numerical Analysis
  \textbf{30} (1993), 1622--1663.

\bibitem{Eyre1998}
David~J. Eyre, \emph{{U}nconditionally gradient stable time marching the
  {C}ahn-{H}illiard equation}, Symposia BB - Computational \& Mathematical
  Models of Microstructural Evolution, MRS Proceedings, vol. 529, 1998,
  pp.~39--46.

\bibitem{Feng2006}
Xiaobing Feng, \emph{{F}ully discrete finite element approximations of the
  {N}avier-{S}tokes-{C}ahn-{H}illiard diffuse interface model for two-phase
  fluid flows}, SIAM Journal on Numerical Analysis \textbf{44} (2006),
  1049--1072.

\bibitem{Gao2012}
Min Gao and Xiao-Ping Wang, \emph{{A} gradient stable scheme for a phase field
  model for the moving contact line problem}, Journal of Computational Physics
  \textbf{231} (2012), 1372 -- 1386.

\bibitem{Gomez2008}
H\'{e}ctor G\'{o}mez, Victor~M. Calo, Yuri Bazilevs, and Thomas~J.R. Hughes,
  \emph{{I}sogeometric analysis of the {C}ahn-{H}illiard phase-field model},
  Computer Methods in Applied Mechanics and Engineering \textbf{197} (2008),
  4333 -- 4352.

\bibitem{Gomez2011}
H\'{e}ctor G\'{o}mez and Thomas~J.R. Hughes, \emph{{P}rovably unconditionally
  stable, second-order time-accurate, mixed variational methods for phase-field
  models}, Journal of Computational Physics \textbf{230} (2011), 5310 -- 5327.

\bibitem{GuillenGonzalez2014}
Francisco Guill\'{e}n-Gonz\'{a}lez and Giordano Tierra, \emph{Second order
  schemes and time-step adaptivity for {A}llen-{C}ahn and {C}ahn-{H}illiard
  models}, Computers \& Mathematics with Applications \textbf{68} (2014), 821
  -- 846.

\bibitem{Han2015}
Daozhi Han and Xiaoming Wang, \emph{{A} second order in time, uniquely
  solvable, unconditionally stable numerical scheme for
  {C}ahn-{H}illiard-{N}avier-{S}tokes equation}, Journal of Computational
  Physics \textbf{290} (2015), 139 -- 156.

\bibitem{Hua2011}
Jinsong Hua, Ping Lin, Chun Liu, and Qi~Wang, \emph{{E}nergy law preserving
  {C0} finite element schemes for phase field models in two-phase flow
  computations}, Journal of Computational Physics \textbf{230} (2011), 7115 --
  7131.

\bibitem{Hyon2010}
Y.~Hyon, D.O.Y. Kwak, and C.~Liu, \emph{{E}nergetic variational approach in
  complex fluids: {M}aximum dissipation principle}, Discrete and Continuous
  Dynamical Systems \textbf{26} (2010), 1291--1304.

\bibitem{Kim2004}
Junseok Kim, Kyungkeun Kang, and John Lowengrub, \emph{{C}onservative multigrid
  methods for {C}ahn-{H}illiard fluids}, Journal of Computational Physics
  \textbf{193} (2004), 511 -- 543.

\bibitem{Kim2004a}
\bysame, \emph{{C}onservative multigrid methods for ternary {C}ahn-{H}illiard
  systems}, Communications in Mathematical Sciences \textbf{2} (2004), 53--77.

\bibitem{Lin2007}
Ping Lin, Chun Liu, and Hui Zhang, \emph{{A}n energy law preserving {C0} finite
  element scheme for simulating the kinematic effects in liquid crystal
  dynamics}, Journal of Computational Physics \textbf{227} (2007), 1411 --
  1427.

\bibitem{Rudraraju2014}
S.~Rudraraju, A.~Van der Ven, and K.~Garikipati, \emph{{T}hree-dimensional
  isogeometric solutions to general boundary value problems of {T}oupin's
  gradient elasticity theory at finite strains}, Computer Methods in Applied
  Mechanics and Engineering \textbf{278} (2014), 705 -- 728.

\bibitem{Rudrarajuetal2015}
S.~Rudraraju, A.~Van~der Ven, and K.~Garikipati, \emph{Mechano-chemical
  spinodal decomposition: {A} phenomenological theory of phase transformations
  in multi-component crystalline solids}, arXiv:1508.05930 (2015), in review.

\bibitem{mechanoChemistryStabilityAnalysisCode}
K.~{Sagiyama}, \emph{{\tt mechanoChemistryStabilityAnalysisCode}: {A library of
  unconditionally stable, second-order accurate schemes for
  mechano-chemistry}},
  \url{https://gitlab.com/compPhysCode/mechanoChemistryStabilityAnalysisCode},
  2015.

\bibitem{Tavakoli2016}
Rouhollah Tavakoli, \emph{{U}nconditionally energy stable time stepping scheme
  for {C}ahn-{M}orral equation: {A}pplication to multi-component spinodal
  decomposition and optimal space tiling}, Journal of Computational Physics
  \textbf{304} (2016), 441 -- 464.

\bibitem{Toupin1964}
R.A. Toupin, \emph{{T}heories of elasticity with couple-stress}, Archive for
  Rational Mechanics and Analysis \textbf{17} (1964), 85--112.

\bibitem{Calo2014}
P.~{Vignal}, L.~{Dalcin}, D.~L. {Brown}, N.~{Collier}, and V.~M. {Calo},
  \emph{{{A}n energy-stable convex splitting for the phase-field crystal
  equation}}, arXiv:1405.3488 (2014).

\bibitem{Wellsetal2006}
G.N. Wells, E.~Kuhl, and K.~Garikipati, \emph{{A} {D}iscontinuous {G}alerkin
  method for the {C}ahn-{H}illiard equation}, Journal of Computational Physics
  \textbf{218} (2006), 860 -- 877.

\bibitem{Wise2010}
S.~M. Wise, \emph{{U}nconditionally stable finite difference, nonlinear
  multigrid simulation of the {C}ahn-{H}illiard-{H}ele-{S}haw system of
  equations}, Journal of Scientific Computing \textbf{44} (2010), 38--68.

\bibitem{Wise2009}
S.~M. Wise, C.~Wang, and J.~S. Lowengrub, \emph{{A}n energy-stable and
  convergent finite-difference scheme for the phase field crystal equation},
  SIAM Journal on Numerical Analysis \textbf{47} (2009), 2269--2288.

\bibitem{Wu2014}
X.~Wu, G.~J. van Zwieten, and K.~G. van~der Zee, \emph{{S}tabilized
  second-order convex splitting schemes for {C}ahn-{H}illiard models with
  application to diffuse-interface tumor-growth models}, International Journal
  for Numerical Methods in Biomedical Engineering \textbf{30} (2014), 180--203.

\end{thebibliography}

\appendix

\end{document}